\documentclass[11pt]{article}
\usepackage{amsmath,bm}
\usepackage{amssymb}
\usepackage{geometry}
\usepackage{graphicx}
\usepackage{hyperref}
\usepackage{subfig}
\usepackage{mciteplus}

\def\gtwid{\mathrel{\raise.3ex\hbox{$>$\kern-.75em\lower1ex\hbox{$\sim
$}}}}
\def\vio{\mathrel{\hbox{$E$\kern-.60em\hbox{$/
$}}}}
\newcommand{\newc}{\newcommand*}

\long\def\begincomment#1\endcomment{%
        \begingroup\sf\baselineskip12pt#1\endgroup}

\newc{\etal}{\textrm{et al.}} 
\newc{\eg}{\textrm{e.g.}} 
\newc{\ie}{\textrm{i.e.}}
\newc{\etc}{\textrm{etc}}
\newc\vs{\textrm{vs.}}
\newc{\cl}{\rm {C.L.}}

\newc{\ev}{\ensuremath{\,\mathrm{eV}}}
\newc{\kev}{\ensuremath{\,\mathrm{keV}}}
\newc{\mev}{\ensuremath{\,\mathrm{MeV}}}
\newc{\gev}{\ensuremath{\,\mathrm{GeV}}}
\newc{\tev}{\ensuremath{\,\mathrm{TeV}}}
\newc{\MeV}{\mev} 
\newc{\TeV}{\tev}
\newc{\invpb}{\ensuremath{/\text{pb}}}
\newc{\invfb}{\ensuremath{/\text{fb}}}
\newc\nb{\ensuremath{\,\mathrm{nb}}} \newc\pb{\ensuremath{\,\mathrm{pb}}} \newc\fb{\ensuremath{\,\mathrm{fb}}}
\newc\pc{\ensuremath{\,\mathrm{pc}}}
\newc\kpc{\ensuremath{\,\mathrm{kpc}}}
\newc\mpc{\ensuremath{\,\mathrm{Mpc}}}
\newc\ps{\ensuremath{\,\mathrm{ps}}} 
\newc\cmeter{\ensuremath{\,\mathrm{cm}}} 
\newc\meter{\ensuremath{\,\mathrm{m}}} 
\newc\kmeter{\ensuremath{\,\mathrm{km}}}
\newc\second{\ensuremath{\,\mathrm{s}}}
\newc\msecond{\ensuremath{\,\mathrm{ms}}}
\newc\nsecond{\ensuremath{\,\mathrm{ns}}}
\newc\psecond{\ensuremath{\,\mathrm{ps}}}

\newc{\chisqmin}{\ensuremath{\chi^2_{\mathrm{min}}}}
\newc{\Delchisq}{\ensuremath{\Delta\chi^2}}
\newc{\chisq}{\ensuremath{\chi^2}}
\newc{\like}{\ensuremath{\mathcal{L}}}

\newc\lsim{\ensuremath{\mathrel{\rlap{\lower4pt\hbox{\hskip1pt$\sim$}}\raise1pt\hbox{$<$}}}}
\newc\gsim{\ensuremath{\mathrel{\rlap{\lower4pt\hbox{\hskip1pt$\sim$}}\raise1pt\hbox{$>$}}}}
\newc{\VEV}[1]{\ensuremath{\langle #1 \rangle}}
\newc{\dl}{\ensuremath{\stackrel{\leftarrow}{D}}}
\newc{\dr}{\ensuremath{\stackrel{\rightarrow}{D}}}

\newc{\bcenter}{\begin{center}}    \newc{\ecenter}{\end{center}}
\newc{\bfl}{\begin{flushleft}}    \newc{\efl}{\end{flushleft}}
\newc{\bfr}{\begin{flushright}}    \newc{\efr}{\end{flushright}}

\newc{\bi}{\begin{itemize}}
\newc{\ei}{\end{itemize}}
\newc{\bed}{\begin{description}}
\newc{\eed}{\end{description}}
\newc{\ben}{\begin{enumerate}}
\newc{\een}{\end{enumerate}}

\newc{\be}{\begin{equation}}
\newc{\ee}{\end{equation}}
\newc{\bea}{\begin{eqnarray}}
\newc{\eea}{\end{eqnarray}}
\newc{\ra}{\rightarrow}

\newc{\alphas}{\ensuremath{\alpha_s}}
\newc{\alphatwo}{\ensuremath{\alpha_2}}
\newc{\alphaone}{\ensuremath{\alpha_1}}
\newc{\alphai}[1]{\ensuremath{\alpha_{#1}}}
\newc{\alphaem}{\ensuremath{\alpha_{\mathrm{em}}}}
\newc{\alphaeff}{\ensuremath{\alpha_{\mathrm{eff}}}}
\newc{\sineff}{\ensuremath{\sin \theta_{\mathrm{eff}}}}
\newc{\sinsqeff}{\ensuremath{\sin^2 \theta_{\mathrm{eff}}}}
\newc{\dalphahad}{\ensuremath{\Delta \alpha_{\mathrm{had}}}}
\newc{\yt}{\ensuremath{h_t}} \newc{\yb}{\ensuremath{h_b}} \newc{\ytau}{\ensuremath{h_{\tau}}}
\newc\mz{\ensuremath{M_Z}} 
\newc\mw{\ensuremath{m_W}}
\newc\mZ{\mz}        \newc\mW{\mw}
\newc\mhsm{\ensuremath{ m_{H_{\mathrm{SM}}}}}
\newc{\mtop}{\ensuremath{ m_t}}               \newc{\mtpole}{\ensuremath{ M_t}}
\newc{\mbottom}{\ensuremath{ m_b}} 
\newc{\mtau}{\ensuremath{ m_{\tau}}}
\newc{\mt}{\mtpole}
\newc{\mb}{\mbottom} 
\newc{\rgg}{\ensuremath{R_{h}(\gamma\gamma)}}
\newc{\rzz}{\ensuremath{R_{h}(ZZ)}}
\newc{\rtwogg}{\ensuremath{R_{h_2}(\gamma\gamma)}}
\newc{\rtwozz}{\ensuremath{R_{h_2}(ZZ)}}
\newc{\ronegg}{\ensuremath{R_{h_1}(\gamma\gamma)}}
\newc{\ronezz}{\ensuremath{R_{h_1}(ZZ)}}
\newc{\rsiggg}{\ensuremath{R_{h_\textrm{sig}}(\gamma\gamma)}}
\newc{\rsigzz}{\ensuremath{R_{h_\textrm{sig}}(ZZ)}}
\newc{\llbar}{\ensuremath{\ell\bar{\ell}}}
\newc{\tauptaum}{\ensuremath{ \tau^+\tau^-}}
\newc{\qqbar}{\ensuremath{ q\bar{q}}} \newc{\ppbar}{\ensuremath{ p\bar{p}}}
\newc{\bbbar}{\ensuremath{ b\bar{b}}} \newc{\ttbar}{\ensuremath{ t\bar{t}}}
\newc{\ffbar}{\ensuremath{ f\bar{f}}} \newc{\tautaubar}{\ensuremath{ \tau\bar{\tau}}}
\newc{\mchi}{\ensuremath{m_\neutone}}
\newc{\squark}{\ensuremath{\tilde{q}}}
\newc{\slepton}{\ensuremath{\tilde{l}}}
\newc{\gluino}{\ensuremath{\tilde{g}}} 
\newc{\mgluino}{\ensuremath{{m_{\gluino}}}}
\newc{\tone}{\ensuremath{{\tilde{t}_1}}}

\newc{\sthw}{\ensuremath{ \sin\theta_W}}              \newc{\cthw}{\ensuremath{\cos\theta_W}}
\newc{\tanthw}{\ensuremath{ \tan\theta_W}}              \newc{\cotthw}{\ensuremath{\cot\theta_W}}
\newc{\ssqthw}{\ensuremath{\sin^2 \theta_W}}
\newc{\msbar}{\ensuremath{\overline{MS}}} \newc{\drbar}{\ensuremath{\overline{DR}}}
\newc{\mtmtsmmsbar}{\ensuremath{ m_t(m_t)^{\msbar}_{{\mathrm{SM}}}}}
\newc{\mtmtsmdrbar}{\ensuremath{ m_t(m_t)^{\drbar}_{{\mathrm{SM}}}}}
\newc{\mtmtmssmdrbar}{\ensuremath{ m_t(m_t)^{\drbar}_{{\mathrm{SUSY}}}}}
\newc{\mbmbmsbar}{\ensuremath{ m_b(m_b)^{\msbar} }}
\newc{\mbmbsmmsbar}{\ensuremath{ m_b(m_b)^{\msbar}_{{\mathrm{SM}}}}}
\newc{\mbmzsmmsbar}{\ensuremath{ m_b(\mz)^{\msbar}_{{\mathrm{SM}}}}}
\newc{\mbmzsmdrbar}{\ensuremath{ m_b(\mz)^{\drbar}_{{\mathrm{SM}}}}}
\newc{\mbmzmssmdrbar}{\ensuremath{ m_b(\mz)^{\drbar}_{{\mathrm{SUSY}}}}}
\newc{\mtaumzsmmsbar}{\ensuremath{ m_{\tau}(\mz)^{\msbar}_{{\mathrm{SM}}}}}
\newc{\mtaumzsmdrbar}{\ensuremath{ m_{\tau}(\mz)^{\drbar}_{{\mathrm{SM}}}}}
\newc{\mtaumzmssmdrbar}{\ensuremath{ m_{\tau}(\mz)^{\drbar}_{{\mathrm{SUSY}}}}}
\newc{\alphasmzms}{\ensuremath{\alpha_s(M_Z)^{\overline{MS}}}}
\newc{\alphaimzms}[1]{\ensuremath{\alpha_{#1}(M_Z)^{\overline{MS}}}}
\newc{\alphaemmz}{\ensuremath{\alpha_{\mathrm{em}}(M_Z)^{\overline{MS}}}}

\newc{\mzero}{\ensuremath{{m_0}}}
\newc{\mhalf}{\ensuremath{ m_{1/2}}}
\newc{\tanb}{\ensuremath{\tan\beta}}
\newc{\azero}{\ensuremath{ A_0}}
\newc{\signmu}{\ensuremath{\rm{sgn}\,\mu}}
\newc{\atau}{\ensuremath{{A_{\tau}}}}
\newc{\mueff}{\ensuremath{\mu_{\rm{eff}}}}
\newc{\lam}{\ensuremath{{\lambda}}}
\newc{\kap}{\ensuremath{{\kappa}}}
\newc{\alam}{\ensuremath{{A_{\lambda}}}}
\newc{\akap}{\ensuremath{{A_{\kappa}}}}
\newc{\hs}{\ensuremath{ H_s}}      
\newc{\mhs}{\ensuremath{ m_{H_s}}} 
\newc{\mgut}{\ensuremath{ M_{\rm GUT}}}
\newc{\mplanck}{\ensuremath{ M_{\rm P}}}      \newc{\mpl}{\ensuremath{ M_{\rm Pl}}}
\newc{\msusy}{\ensuremath{ M_{\rm SUSY}}}      \newc{\ms}{\ensuremath{ M_{\rm S}}}
 \newc{\hu}{\ensuremath{ H_u}}       \newc{\hd}{\ensuremath{ H_d}}
 \newc{\mhu}{\ensuremath{ m_{H_u}}}       \newc{\mhd}{\ensuremath{ m_{H_d}}}
 \newc{\mhuew}{\ensuremath{ m^{\ast}_{H_u}}}       \newc{\mhdew}{\ensuremath{ m^{\ast}_{H_d}}}
 \newc{\mhuewsq}{\ensuremath{ m^{\ast\, 2}_{H_u}}}       \newc{\mhdewsq}{\ensuremath{ m^{\ast\, 2}_{H_d}}}
 \newc{\mhl}{\ensuremath{m_\hl}} 
 \newc{\mhone}{\ensuremath{m_{h_1}}} 
 \newc{\mhtwo}{\ensuremath{m_{h_2}}} 
 \newc{\mglu}{\ensuremath{m_{\tilde g}}} 
 \newc{\mul}{\ensuremath{m_{\tilde{u}_L}}} 
 \newc{\mtone}{\ensuremath{m_{\tilde{t}_1}}} 
 \newc{\ma}{\ensuremath{m_A}} 
 \newc{\maone}{\ensuremath{m_{a_1}}} 
 \newc{\matwo}{\ensuremath{m_{a_2}}}
 \newc{\hone}{\ensuremath{h_1}}
 \newc{\htwo}{\ensuremath{h_2}}
 \newc{\aone}{\ensuremath{a_1}}
 \newc{\atwo}{\ensuremath{a_2}}
 \newc{\mqthree}{\ensuremath{m_{\tilde{Q}_3}^2}}
 \newc{\muthree}{\ensuremath{m_{\tilde{u}_3}^2}}

\newc{\sigsip}{\ensuremath{\sigma^{\rm SI}_{p}}}	\newc{\sigsin}{\ensuremath{\sigma^{\rm SI}_{n}}}
\newc{\sigsdp}{\ensuremath{\sigma^{\rm SD}_{p}}}	\newc{\sigsdn}{\ensuremath{\sigma^{\rm SD}_{n}}}
\newc{\sigsi}{\ensuremath{\sigma^{\rm SI}}}	\newc{\sigsd}{\ensuremath{\sigma^{\rm SD}}}
\newc{\abund}{\ensuremath{ \Omega h^2}}
\newc{\omegadm}{\ensuremath{ \Omega_{{\rm DM}}}}     \newc{\abunddm}{\ensuremath{ \Omega_{{\rm DM}} h^2}} 
\newc{\omegam}{\ensuremath{ \Omega_{{\rm m}}}}       \newc{\abundm}{\ensuremath{ \Omega_{{\rm m}} h^2}}
\newc{\omegab}{\ensuremath{ \Omega_{{\rm b}}}}	\newc{\abundb}{\ensuremath{ \Omega_{{\rm b}} h^2}}
\newc{\omegatot}{\ensuremath{ \Omega_{{\rm TOT}}}}
\newc{\omegacdm}{\ensuremath{ \Omega_{{\rm CDM}}}}   \newc{\abundcdm}{\ensuremath{ \Omega_{{\rm CDM}} h^2}}
\newc{\omegalambda}{\ensuremath{ \Omega_{\Lambda}}} \newc{\abundlambda}{\ensuremath{ \Omega_{\Lambda} h^2}}
\newc{\omegarad}{\ensuremath{ \Omega_{{\rm rad}}}}  \newc{\abundrad}{\ensuremath{ \Omega_{{\rm rad}} h^2}}
\newc{\rhocrit}{\ensuremath{ \rho_{\rm crit}}}
\newc{\rhochi}{\ensuremath{ \rho_{\chi}}}
\newc{\abunchi}{\ensuremath{\Omega_\chi h^2}}
\newc{\abundlsp}{\ensuremath{\Omega_{\rm LSP}h^2}}

\newc{\amu}{\ensuremath{ a_{\mu}}}        \newc{\amususy}{\ensuremath{ a_{\mu}^{\mathrm{SUSY}}}}
\newc{\amuexpt}{\ensuremath{ a_{\mu}^{\mathrm{expt}}}}        \newc{\amusm}{\ensuremath{ a_{\mu}^{\mathrm{SM}}}}
\newc\deltaamu{\ensuremath{\Delta a_{\mu}}} \newc{\deltaamususy}{\ensuremath{\delta a_{\mu}^{\mathrm{SUSY}}}}
\newc\gmtwo{\ensuremath{ (g-2)_{\mu}}} 
\newc{\deltagmtwomususy}{\ensuremath{\delta\left(g-2\right)_{\mu}^{\mathrm{SUSY}}}}
\newc{\deltagmtwomu}{\ensuremath{\delta\left(g-2\right)_{\mu}}}
\newc\BR{\ensuremath{\rm BR}}
\newc\bsgamma{\ensuremath{ b\rightarrow s \gamma }}
\newc\bxsgamma{\ensuremath{\overline{B}\rightarrow X_{s}\gamma}}
\newc\brbsgamma{\ensuremath{\BR\left(\bsgamma\right)}}
\newc\brbxsgamma{\ensuremath{\BR\left(\bxsgamma\right)}}
\newc\bsmumu{\ensuremath{B_s\to\mu^+\mu^-}}
\newc\brbsmumu{\ensuremath{\BR\left(B_s\to\mu^+\mu^-\right)}}
\newc\bdmmumu{\ensuremath{\overline{B}_d\to\mu^+\mu^-}}
\newc\bbbarmix{\ensuremath{\overline{B}_s\mbox{-}B_s}}      
\newc\delmbs{\ensuremath{\Delta M_{B_s}}}
\newc{\butaunu}{\ensuremath{B_u \rightarrow \tau \nu}}
\newc{\brbutaunu}{\ensuremath{\BR\left(B_u \rightarrow \tau \nu\right)}}

\newcommand*{\reffig}[1]{Fig.~\ref{#1}}

     \newcommand*{\refsec}[1]{Sec.~\ref{#1}}

\newcommand*{\neutone}{\ensuremath{\tilde{\chi}^0_1}}
\newcommand*{\neuttwo}{\ensuremath{\tilde{{\chi}}^0_2}}

\newcommand*{\charone}{\ensuremath{\tilde{{\chi}}^{\pm}_1}}

\newcommand*{\slep}{\ensuremath{\tilde{l}}}

\newcommand*{\seven}{\ensuremath{\sqrt{s}=7\tev}}
\newcommand*{\eight}{\ensuremath{\sqrt{s}=8\tev}}
\newcommand*{\alphaT}{\ensuremath{\alpha_T}}

\let\oldcite\cite
\renewcommand*{\cite}{~\oldcite}

\newcommand*{\hl}{\ensuremath{h}}

\begin{document}

\title{Natural MSSM after the LHC 8 TeV run}  
\author{Kamila Kowalska and Enrico Maria Sessolo\\[2ex]
\small\it National Centre for Nuclear Research, Ho$\dot{z}$a 69, 00-681 Warsaw, Poland\\
}
\date{}
\maketitle
\centering
\url{Kamila.Kowalska@fuw.edu.pl, Enrico-Maria.Sessolo@fuw.edu.pl}

\abstract{We investigate the impact of direct LHC SUSY searches on the parameter space of three natural scenarios in the MSSM.  
In the first case the spectrum consists of light stops, sbottoms, and Higgsino-like neutralinos, 
while the other particles are assumed to be out of the experimental reach. In the second case we consider an additional light gluino.  
Finally we study a more complex spectrum comprising also light sleptons, wino-like chargino, and a bino-like neutralino. We simulate 
in detail three LHC searches: stop production at ATLAS with 20.7/fb, CMS 11.7/fb 
inclusive search for squarks and gluinos with the variable \alphaT, and CMS 9.2/fb electroweak production with 3 leptons in the final state. 
For each point in our scans we calculate the exclusion likelihood due to the individual searches and to their statistical combination.
We calculate the fine-tuning measure of the points allowed by the LHC and the implications for the Higgs mass 
and other phenomenological observables: Higgs signal rates, the relic density, \brbsmumu, \brbxsgamma, and the spin-independent neutralino-proton scattering cross section. 
We find that points with acceptable levels of fine-tuning are for the most part already excluded by the LHC 
and including the other constraints further reduces the overall naturalness of our scenarios.}

\newpage

\section{Introduction}\label{intro:sec}

With the end of 2012, the LHC completed its \eight\ run, 
and both the ATLAS and CMS Collaborations collected approximately 21\invfb\ of data.
Many physics analyses have been already completed and made public by the two collaborations, 
in the framework of the Standard Model (SM) and beyond (BSM).
Additional analyses are scheduled to appear in the next few months. 
Undoubtedly, the greatest success has been the observation of 
the Higgs boson of the SM\cite{Chatrchyan:2012ufa,Aad:2012tfa}, or at least of a particle that couples to the SM with very similar strength,
with mass $\mhl\simeq125\gev$. On the other hand, direct searches for new BSM physics,
which in the largest share are designed for the observation of low energy supersymmetry (SUSY),
have given null results to this point.

In the context of SUSY the latest LHC results just mentioned 
(the discovery of the Higgs boson, the nonobservation  
of light SUSY particles, but also the first evidence of a SM-like \brbsmumu\ at LHCb\cite{Aaij:2012nna,Aaij:2013aka} and CMS\cite{Chatrchyan:2013bka}) 
seem to point to the fact that within the framework of the Minimal Supersymmetric Standard Model (MSSM) 
the typical scale of the superpartners, defined as the geometric mean of the stop masses, $\msusy=(m_{\tilde{t}_1}m_{\tilde{t}_2})^{1/2}$, 
is higher than the scale presently testable with direct searches. In fact,
in the MSSM, $\mhl\simeq125\gev$
requires stops in the multi-TeV regime, unless one accounts for nearly maximal stop mixing, 
$|X_t|/\msusy\simeq\sqrt{6}$\cite{Dermisek:2007fi,*Low:2009nj,Hall:2011aa,Heinemeyer:2011aa,Baer:2011ab,*Arbey:2011ab,
*Carena:2011aa,*Cao:2012fz,*Christensen:2012ei,*Brummer:2012ns,*Arbey:2012dq,CahillRowley:2012rv}.
While this fact does not pose any particular problem
from the phenomenological point of view (see, e.g.,\cite{Fowlie:2013oua} for a recent global analysis),  
the implications of large \msusy\ have exacerbated the ``naturalness'' problem of the MSSM,
also called in the literature the ``Little Hierarchy" problem\cite{Ellis:1986yg,*Barbieri:1987fn,Ross:1992tz,*Carlos:1993yy,*Anderson:1994dz,*Anderson:1994tr,*Dimopoulos:1995mi,*Ciafaloni:1996zh,
*Bhattacharyya:1996dw,*Chankowski:1997zh,*Chan:1997bi,*Barbieri:1998uv,*Wright:1998mk,*Chankowski:1998xv,*Kane:1998im,*Giusti:1998gz,
*Feng:1999zg,*Chacko:2005ra,*Choi:2005hd,*Kitano:2005wc,*Perelstein:2007nx,*Essig:2007kh,*Cassel:2009ps,*Barbieri:2009ev,
Gogoladze:2009bd,*Horton:2009ed,*Lodone:2010kt}, i.e., the requirement
that the electroweak (EW) scale be obtained without excessive fine-tuning of the soft SUSY-breaking 
terms in the Lagrangian.

To put the issue in more quantitative terms,
let us consider a measurement of fine-tuning for the soft SUSY-breaking terms (here generically indicated with $p_i$) that enter the minimization
conditions of the scalar potential:   
for instance the well known Barbieri-Giudice measure\cite{Ellis:1986yg,*Barbieri:1987fn}, $\Delta=\max \{\Delta_{p_i}\}$, with
\be 
\Delta_{p_i}=\left|\frac{\partial\log M_Z^2}{\partial\log p_i^2}\right|\,.
\label{finetune}
\ee
One can calculate $\Delta$ (by using, e.g., the formulas of\cite{Papucci:2011wy}) 
for the values of the soft terms that are favored at $2\sigma$ by the Higgs mass measurement. 
The obtained fine-tuning depends on the scale of the SUSY-breaking sector, $\Lambda$:
if $\Lambda=10\tev$ one gets $\Delta\sim40-100$ for $m_{\tilde{t}_1},m_{\tilde{t}_2}\sim 600-1000\gev$ 
and maximal stop mixing (provided $\mu$ does not exceed $500-600\gev$), and $\Delta\simeq200$ or more 
for $m_{\tilde{t}_1},m_{\tilde{t}_2}>3000\gev$ with zero mixing; if, on the other hand, $\Lambda\sim10^{16}\gev$, then $\Delta$ increases by an order of magnitude or more,
depending on the value of the gluino mass (although for very large $\Lambda$ the leading log (LL)
approximation must be taken with caution\cite{Baer:2012mv,*Baer:2013bba,Hardy:2013ywa}).
 
Thus, in the MSSM the measured value of the Higgs mass 
requires a large amount of fine-tuning. (Addition of extra sectors can ameliorate this problem by raising the value of the tree-level Higgs mass, like in the case of the Next-to-Minimal Supersymmetric SM\cite{BasteroGil:2000bw,*Delgado:2010uj,*Ellwanger:2011mu,*Ross:2011xv}, or in interesting alternatives like\cite{Athron:2013ipa}.)
On the other hand, since $\Delta$ is generally larger in the case with multi-TeV stop masses than in the case where the 
correct Higgs mass is obtained thanks to maximal stop mixing, the idea of Natural SUSY, which finds its origin in 
many of the papers cited in Ref.\cite{Ross:1992tz} and also includes the concept of Effective SUSY\cite{Cohen:1996vb}, has seen a revival in the last couple of years\cite{Hall:2011aa,Papucci:2011wy,Cheung:2005pv,*Baer:2011ec,Kitano:2006gv,*Asano:2010ut,*Cassel:2011tg,*Sakurai:2011pt,*Perelstein:2011tg,
*Brust:2011tb,*Baer:2012uy,*Baer:2012up,*Baer:2012cf,*Lodone:2012kp,*Lee:2012sy,*Perelstein:2012qg,*Cao:2012rz,*Grothaus:2012js,*Mescia:2012fg,
*Gogoladze:2012yf,*Boehm:2013qva,*Gogoladze:2013wva,
Blanke:2013uia,Buchmueller:2013exa,Kribs:2013lua}.
In fact, Natural SUSY spectra are characterized by the presence of light stops and sbottoms --which are not as much constrained by the LHC searches as the first two 
generations' squarks --by a small value of the $\mu$ parameter, and by heavy masses for the remaining squarks.
Interestingly, ATLAS and CMS have followed this lead 
and their interpretations of the results from direct SUSY searches  
have shifted from being heavily oriented towards
constrained models like the Constrained MSSM\cite{Kane:1993td}, to simplified models (SMS)\cite{Chatrchyan:2013sza} designed to exclude particles 
more in line with the naturalness requirement.

The interpretations of SMS bounds
on the production cross section times branching ratio (BR) for particular signal topologies
give a good approximation and a qualitative picture useful for drawing conclusions even in more complex models.
Nonetheless, their accuracy in reproducing the exclusion limits that would be obtained in a more generic scenario depends strongly on the 
relative magnitude of the BR and experimental efficiencies in the selected topology with respect to other possible final states.
Although this is rarely a problem for Natural SUSY scenarios, characterized by a limited number of light particles,
it also is not difficult to imagine possible models in which the decay BR to final states for which
the selected search has little sensitivity is dominant. 
On the other hand, these problems can be avoided by simulating in detail 
the experimental searches with a likelihood function approach, as was recently done 
in\cite{Fowlie:2013oua}, where the statistical impact of two LHC SUSY searches, 
the CMS \alphaT\ search with 11.7\invfb\ integrated luminosity\cite{Chatrchyan:2013lya} and the 
CMS \textit{3-lepton} search for EW production\cite{CMS-PAS-SUS-12-022}, 
was calculated on the parameter space of a 9-dimensional parametrization of the MSSM.
Moreover, calculation of a likelihood function allows one to statistically combine 
limits from different independent searches on the parameter space of the analyzed model.  

Note also that a detailed simulation of an LHC search for a complex model 
can produce limits on a certain particle's mass that are stronger than the ones obtained in a SMS 
involving the same particle. This could be due to the presence of 
two (or more) particles producing indistinguishable signatures at the detector level,
as recently shown in\cite{Kribs:2013lua}, where a LHC analysis of 
Natural SUSY-type of spectra involving light Higgsinos, $\tilde{t}_L$,  $\tilde{b}_L$,  and $\tilde{t}_R$  
was performed.
Or, if all available production channels are open, additional limits on the mass of a certain particle can be put indirectly 
by the production and decay of a different particle if the spectra show some correlation. 
This issue was discussed in the context of bounds on third generation squarks and gluinos in\cite{Buchmueller:2013exa}, 
where the necessity of combining different experimental signatures was also emphasized.  


In this paper, following the procedure for the implementation of LHC SUSY searches adopted in\cite{Fowlie:2013oua}, 
we perform a similar analysis for 
three MSSM scenarios, whose spectra are natural in the sense described by Eq.~(\ref{finetune}).
We consider the following cases, ordered with increasing complexity in the spectrum: 
1. The spectrum consists of light $\tilde{t}_1$,  $\tilde{b}_1$,  $\tilde{t}_2$,  
and Higgsino-like neutralinos;
2. The spectrum includes also light gluinos; 3. 
The spectrum consists of the same particles as in Scenario 2, with the exceptions that the lightest
neutralino is bino-like, the lightest chargino is wino-like, and there are light sleptons. 

For each scenario we generate a random sample of points. 
For each point we perform on-the-fly simulation of the LHC signal, from generation of the hard scattering  
events to simulation of the detector's response to calculate the efficiencies 
(see also\cite{Fowlie:2011mb,Fowlie:2012im,Kowalska:2012gs,Kowalska:2013hha} for a description of this procedure), 
and compare the signal to the observed and background yields, provided by the experimental collaborations,
through construction of a likelihood function. 
We consider three LHC searches based on the \eight\ data set: 21\invfb\ ATLAS direct stop production
with 1 lepton in the final state\cite{ATLAS-CONF-2013-037} and the two searches that were already used in\cite{Fowlie:2013oua}:
9.2\invfb\ CMS 3-lepton EW production and 11.7\invfb\ CMS \alphaT\ inclusive search. 
However, we updated the procedure of\cite{Fowlie:2013oua} by including
the next-to-leading-order and next-to-leading-log (NLO+NLL) corrections to the production cross sections. 
We then consider statistical combinations of the implemented searches for our three scenarios
and derive combined limits on the sparticle masses. 
This is similar in spirit 
to the procedure adopted in\cite{Buchmueller:2013exa}, which   
used some of the CMS searches from the \seven\ data set. 
Finally, for the points in our scenarios that are not excluded at the 95\%~C.L. we calculate the fine-tuning measure 
according to Eq.~(\ref{finetune}), as well as some relevant phenomenological observables:
Higgs mass and signal rates, relic density, \brbxsgamma, \brbsmumu, and the spin-independent (SI) neutralino-proton scattering cross section
\sigsip.

We limit ourselves to regions of the parameter space over which the LHC searches we simulate have 
significant sensitivity. This means that we do not treat here the case of compressed spectra, 
for which $|m_{\tilde{t},\tilde{b},\tilde{g}}-m_{\neutone}|/m_{\tilde{t},\tilde{b},\tilde{g}}\ll1$. 
It is known that those regions are ``pockets" in which Natural SUSY could be hiding\cite{Kribs:2013lua}.

Our analysis presents elements in common with the works mentioned above, Refs.\cite{Buchmueller:2013exa} and\cite{Kribs:2013lua}, 
but we also show several novel features: (i) The LHC searches we select involve third generation squarks, gluinos, and EW-produced
charginos and neutralinos, and they are all based on the \eight\ data set. (ii) We consider very general, $R$-parity conserving, 
loosely natural MSSM spectra to analyze some interesting effects  
(limits from EW production, decays of gluinos and third generation squarks through off- and on-shell sleptons).
(iii) We quantify the fine-tuning for all our points and analyze the
impact of phenomenological constraints other than the direct searches at the LHC.  

This paper is organized as follows. 
In \refsec{nat:sec} we summarize the features of natural MSSM spectra and we define the three 
scenarios considered in this analysis. 
In \refsec{lhc:sec} we describe our procedure for deriving 
the likelihood functions for direct SUSY searches at the LHC 
and we present the results of their validation against the official limits from ATLAS and CMS. 
Section~\ref{res:sec} is devoted to the discussion of the results.
We summarize our findings in \refsec{sum:sec}.

\section{Naturalness in the MSSM}\label{nat:sec}

The concept of Natural SUSY is closely related to the EW symmetry-breaking 
mechanism and has been widely discussed in the literature. 
Here we briefly recall its most important features.

One of the minimization conditions of the scalar potential allows one to express the mass of the $Z$ boson 
in terms of the running soft terms \mhu, \mhd\ and $\mu$:
\be
\frac{1}{2}M_Z^2=-\mu^2+\frac{(\mhd^2+\Sigma_d)-(\mhu^2+\Sigma_u)\tan^2\beta}{\tan^2\beta-1}\,,
\label{mz}
\ee
where $\Sigma_u$ and $\Sigma_d$ are the radiative corrections to the tree-level potential, 
which depend on the running of SUSY parameters. For moderate to large \tanb\ ($\tanb >8$), 
the \mhd\ term can be neglected and the correct value of $M_Z$ is obtained through 
the cancellation between the $\mu^2$, $\mhu^2$ and $\Sigma_u$ terms. 
The naturalness criterion\cite{Ellis:1986yg,*Barbieri:1987fn} states that $\mu^2$ and $\Sigma_u$ should be 
of the order of the EW symmetry-breaking scale (squared) 
in order to avoid excessive, or ``unnatural," fine-tuning of the model parameters. 

A widely used measure of the EW fine-tuning associated with 
the parameters of the model is given in Eq.~(\ref{finetune}).
The total measure $\Delta$ for a given model point is the
maximal contribution to the fine-tuning among all of the model's parameters.
A precise determination of the amount of fine-tuning that makes a model unnatural is 
somewhat a matter of taste. In the literature it is usually assumed that a viable amount is $\Delta^{-1}\sim10\%-20\%$. 

In this paper we will be more conservative and assume an upper bound for our generated spectra 
$\Delta^{-1}\geq 1\%$, or $\Delta\leq 100$. 
We can easily translate this requirement into upper bounds for the soft terms\cite{Papucci:2011wy}. 
From Eq.~(\ref{mz}) one can see that the $\mu$ parameter cannot exceed $M_Z$ by 1 order of magnitude, 
which implies fairly light Higgsinos. 
By calculating the measure of Eq.~(\ref{finetune}) from Eq.~(\ref{mz}) and imposing $\Delta\leq 100$ one gets
\be
|\mu|\lesssim 645\gev\,.\label{muFT}
\ee

Secondly, since the dominant loop contribution to $\Sigma_u$ comes from the top Yukawa and the squarks of the third generation, 
and it is given in the LL approximation by\cite{Martin:1997ns,Papucci:2011wy}
\be
\Sigma_u|_{\textrm{stop}}=-\frac{3 y_t^2}{8\pi^2}(\mqthree+\muthree+|A_t|^2)\log\left(\frac{\Lambda}{\tev}\right)\,,\label{3genFT_1}
\ee
imposing $\Delta\leq 100$ places a direct constraint on the third generation soft masses and mixing, 
\be
(\mqthree+\muthree+|A_t|^2)\lesssim (3700 \gev)^2\log\left(\frac{\Lambda}{\tev}\right)^{-1}\,,\label{3genFT_2}
\ee
where $\Lambda$ is the scale at which SUSY breaking is transmitted to the MSSM. 
(The bounds become increasingly more severe when raising $\Lambda$ by orders of magnitude above the TeV scale.)

The one-loop contribution to Eq.~(\ref{mz}) due to a Majorana wino reads
\be
\Sigma_u|_{M_2}=-\frac{3 g_2^2}{8\pi^2}|M_2|^2\log\left(\frac{\Lambda}{\tev}\right)\,,\label{M2FT}
\ee
so that $\Delta\leq 100$ gives
\be
|M_2|\lesssim 5400 \gev\cdot\log\left(\frac{\Lambda}{\tev}\right)^{-1/2}.
\ee

Finally, the contribution from a Majorana gluino to the stop mass can be significant, 
introducing a non-negligible two-loop contribution to the $\Sigma_u$ term,
\be
\Sigma_u|_{M_3}=-\frac{2 y_t^2}{\pi^3}\alpha_s|M_3|^2\log^2\left(\frac{\Lambda}{\tev}\right)\,.\label{M3FT}
\ee
One gets, for the gluino mass parameter $M_3$,\footnote{In the case of Dirac gluinos the limit is weaker\cite{Fox:2002bu,Papucci:2011wy,Hardy:2013ywa}.}
\be
|M_3|\lesssim 8500\gev\cdot\log\left(\frac{\Lambda}{\tev}\right)^{-1}\,.\label{M3FT}
\ee

The other particles in the spectrum can either have a much larger mass (masses of the squarks of the first two generations 
are already pushed well above 1\tev\ by the limits from direct SUSY searches at the LHC)
or are allowed to be at the same mass scale as the light ones. 
Such a possibility is particularly interesting in the case of sleptons, 
since it opens a way of testing a model with direct EW production of charginos and neutralinos. 
On the other hand, allowing different compositions for the lightest neutralino (by assuming $M_1,M_2<\mu$) 
would allow one to investigate different scenarios for generating the dark matter in the Universe.\bigskip

\begin{table}[t]\footnotesize
\begin{centering}
\begin{tabular}{|p{4.5cm}|p{4.5cm}|p{5.0cm}|}
\hline 
Scenario 1 & Scenario 2 & Scenario 3\\
\hline 
$M_1=3\tev$ & $M_1=3\tev$ & $0.01\tev\le M_1\le0.4\tev$\\
 & & $M_1<M_2$\\
$M_2=1.5\tev$  & $M_2=1.5\tev$ &  $0.1\tev\le M_2\le 0.63\tev$\\
$M_3=1.6\tev$ & $0.1\tev\le M_3\le1.6\tev$ & $0.1\tev\le M_3\le1.6\tev$\\
$m_{\tilde{L}_{1,2,3}}=m_{\tilde{e}_1}=m_{\tilde{e}_2}=m_{\tilde{e}_3}=3\tev$ & $m_{\tilde{L}_{1,2,3}}=m_{\tilde{e}_1}=m_{\tilde{e}_2}=m_{\tilde{e}_3}=3\tev$ & $0.1\tev\leq m_{\tilde{L}_{1,2,3}},m_{\tilde{e}_1},m_{\tilde{e}_2},m_{\tilde{e}_3}\leq0.63\tev$ \\
$0.075\tev\le\mu\le0.63\tev$ & $0.075\tev\le\mu\le0.63\tev$ & $\mu=0.63\tev$\\
$0.1\tev\le m_{\tilde{Q}_3},m_{\tilde{u}_3}\le 1.4\tev$ & $0.1\tev\le m_{\tilde{Q}_3},m_{\tilde{u}_3}\le 1.4\tev$ & $0.1\tev\le m_{\tilde{Q}_3},m_{\tilde{u}_3}\le 1.4\tev$\\
\hline
$\tilde{t}_{1,2}$, $\tilde{b}_{1}$, \neutone, \neuttwo, \charone & $\tilde{g}$, $\tilde{t}_{1,2}$, $\tilde{b}_{1}$, \neutone, \neuttwo, \charone &
Sleptons, $\tilde{g}$, $\tilde{t}_{1,2}$, $\tilde{b}_{1}$, \neutone, \neuttwo, \charone\\
\hline
\end{tabular}
\caption{\footnotesize Soft SUSY-breaking parameters characteristic of the natural scenarios considered in this study. 
The bottom line shows the light particles present in each spectrum.}
\label{tab:scenarios}
\end{centering}
\end{table}

As mentioned in \refsec{intro:sec}, we construct three scenarios in the MSSM, 
with characteristic spectra subject to the bounds of Eqs.~(\ref{muFT})-(\ref{M3FT}), for a conservative value
$\Lambda=10\tev$. 
We randomly scan the parameters of the phenomenological MSSM 
(parametrized in its unconstrained version by 24 free parameters defined at \msusy),
on which we impose conditions leading to natural spectra. 
We assume that the squarks of the first two generations are out of reach at the LHC, $m_{\tilde{Q}_{1,2}}=m_{\tilde{u}_1}=m_{\tilde{u}_2}=m_{\tilde{d}_1}=m_{\tilde{d}_2}=5\tev$. 
Similarly, we set $m_{\tilde{d}_3}=5\tev$,
and fix $A_b=A_{\tau}=-0.5\tev$. 
$A_t$, \tanb, and \ma\ are free to vary in the following ranges: 
$-2\tev\le A_t\le 2\tev$, 
$3\le \tanb\le 62$, and
$0.1\tev\le m_A\le 2\tev$, respectively. 
Note that the upper limit on $|A_t|$ is imposed to satisfy $\Delta\leq100$; see Eq.~(\ref{3genFT_2}). 
The scanning ranges of the remaining parameters are summarized for each scenario in Table~\ref{tab:scenarios}. 
Where relevant, we impose LEP limits\cite{Beringer:1900zz} on the masses of charginos, sleptons and neutralinos.
Notice that, given our choices for $M_3$ and $M_2$, the fine-tuning measure associated with those parameters is always  $\Delta_{M_{2,3}}\lesssim 5$ (for $\Lambda\simeq10\tev$), 
so that the main contribution to the total $\Delta$ comes from the third generation squarks and Higgsino sector. 
Notice also that we do not make any additional assumptions about the mass hierarchy between the light sparticles, 
as well as the mixing in the stop sector. We differ in this from\cite{Buchmueller:2013exa} and\cite{Kribs:2013lua}. 
Finally, our choice of gaugino mass parameters in 
Scenario 3 will allow us to investigate the impact of the LHC searches 
in the EW sector.

For each scenario we create a sample of more than 5000 points subject to the following constraints, whose central values are 
taken from Table~2 of Ref.\cite{Fowlie:2013oua} and the uncertainties are obtained from the same table by adding the experimental and theoretical errors in quadrature.
\brbxsgamma\ and \brbsmumu\ are always satisfied at $2\sigma$.\footnote{With respect to Table~2 of Ref.\cite{Fowlie:2013oua}, the measured value of \brbsmumu\ was very recently updated by the LHCb and CMS Collaborations\cite{Aaij:2013aka,Chatrchyan:2013bka}. 
We checked that the vast majority of our points still fall within $2\sigma$ of the new determinations.}
For the relic density we impose only an 
upper limit at $2\sigma$, as it is well known that small Higgsino masses
tend to create an underabundance of present-day dark matter with respect to the central value measured by PLANCK\cite{Ade:2013zuv}
or WMAP\cite{Komatsu:2010fb}. 
This is not 
necessarily a problem for the model, 
since it is easy to conceive plausible mechanisms and additional particles that
can boost the value of the relic density,
as explained, e.g., in\cite{Baer:2013ula} and references therein.

The theoretical uncertainty on the Higgs mass calculation given in Table~2 of Ref.\cite{Fowlie:2013oua} amounts to 
$3\gev$\cite{Heinemeyer:2011aa} and is thus dominant with respect to the experimental uncertainty, 0.6\gev.
We initially require the points in our sample to be consistent with theoretical and experimental uncertainty at $2\sigma$.
Note that given our choice of parameter scanning ranges, driven by $\Delta\leq100$, 
a Higgs mass close to or larger than $125\gev$ becomes very difficult to obtain.
On the other hand, a conservative window of $2\sigma$ around the central value leads to an underpopulation of points in the region that is more interesting for investigating the impact of the LHC, at $\msusy\lesssim1\tev$. 
Since the main focus of this paper is to analyze the impact of 
LHC searches on natural spectra, we extended the initial sample with points characterized by $\msusy\lesssim1\tev$, irrespective of the Higgs mass constraint. We include these points when showing our results in \refsec{res:sec}.
 
Additionally, for all the points we calculated the Higgs signal rates \rgg\ and \rzz. We do not impose, however, 
constraints on those observables when constructing our samples since there is a $2\sigma$ discrepancy between the CMS and ATLAS results in the $\gamma\gamma$ channel\cite{ATLAS-CONF-2013-012,*CMS-PAS-HIG-13-001}. Nevertheless, we comment on the impact of both determinations in \refsec{res:sec}.

The mass spectra are calculated with \texttt{softsusy-3.3.6}\cite{softsusy},
\brbxsgamma\ and \brbsmumu\ with \texttt{superiso v3.3}\cite{superiso},
the relic density and \sigsip\ with \texttt{MicrOMEGAs 2.4.5}\cite{micromegas}.
The Higgs signal rates are computed using \texttt{FeynHggs 2.9.4}\cite{feynhiggs:99,*feynhiggs:00,*feynhiggs:03,*feynhiggs:06} 
based on the procedure described in Sec.~4.3 of Ref.\cite{Fowlie:2013oua}.
The numerical codes are interfaced through the package BayesFITS,
described in detail in\cite{Fowlie:2012im,Kowalska:2012gs,Fowlie:2013oua}.

\section{LHC SUSY limits}\label{lhc:sec}

In this section we describe our implementation of the LHC SUSY limits.
To validate the accuracy of our procedure, we also show here the results of applying the searches to some of the 
SMS designed by the experimental collaborations.

We extend the procedure developed in\cite{Fowlie:2013oua}.
For each implemented search we construct an approximate but accurate likelihood function, which yields an exclusion confidence level 
for each point in our samples. 
The likelihood is obtained through an algorithm that mimics 
the analyses performed by the experimental collaborations. 
For every point in the parameter space we calculate the decay BR with \texttt{SUSYHIT}\cite{susyhit}, 
generate 5000 events at the scattering level with \texttt{PYTHIA6.4}\cite{PYTHIA}, and pass the hadronization products to the fast detector simulator \texttt{PGS4}\cite{PGS4}.
From the physical objects produced by the detector simulator, we construct the kinematical variables, 
\alphaT, $H_T$, $M_T$, $m_{\textrm{eff}}$, $am_{T_2}$, $m_{jjj}$, proper of the three searches considered here (described below) and apply the selection cuts. 
We use the CMS and ATLAS detector cards respectively, with the settings recommended by both collaborations. 
We also tune the $b$-tagging algorithm used by \texttt{PGS4} in order to reproduce the corresponding efficiencies reported by CMS\cite{Chatrchyan:2012paa} and ATLAS\cite{ATLASbtag}. 
This step is particularly important, since $b$ tagging plays a crucial role in deriving the exclusion bounds for the squarks of the third generation. 
Finally, different kinematical bins $i$ are constructed, closely following the experimental papers, 
the cuts are applied and the acceptances/efficiencies $\varepsilon_i$ are calculated as the fraction of events that pass all the cuts. 
We use NLO+NNL cross sections, $\sigma_{\textrm{NLO+NLL}}$, provided by the LHC SUSY Cross Section Working Group\cite{LHCSXSECWG}.

The number of signal events in a given bin is calculated as $s_i=\varepsilon_i\times\sigma_{\textrm{NLO+NLL}}\times\int L$, where $\int L$ 
is the integrated luminosity.
The obtained signal yields are finally statistically compared to the publicly available observed ($o_i$) and background ($b_i$) yields of the searches, 
provided in the experimental papers, as described in\cite{Fowlie:2012im,Kowalska:2013hha}.
The systematic uncertainties on the background yields ($\delta b_i$) are accounted for in our analysis 
by convolving the Poisson distribution $P$ with a Gaussian or log-normal (depending on the bin\cite{Fowlie:2012im}) distribution $G$.
The likelihood function for each bin is thus calculated:
\begin{equation}
\mathcal{L}_i(o_i, s_i, b_i)=\int P(o_i|s_i,\bar{b}_i)G(\bar{b}_i|b_i,\delta b_i)d\bar{b}_i\,,\label{likelihood}
\end{equation}
and the final likelihood for each point is the product of the likelihoods for each separate bin. 
The appropriate confidence level is obtained from the $\delta\chisq$ variable as $\delta\chisq=-2\log(\mathcal{L}/\mathcal{L}_{\textrm{max}})$.

Both ATLAS and CMS performed many analyses at \eight\ with different experimental signatures. 
For the purpose of this paper, we implement the analyses that either present 
the strongest exclusion limits on the mass of a particle under study\footnote{In the days preceding the submission of this paper the CMS Collaboration 
updated the results of the EW search to 19.5\invfb\cite{CMS-PAS-SUS-13-006}. 
While the limits from EW production in Scenario~3 will become even more severe, we do not expect significant qualitative differences for the results presented in \refsec{res:sec}.} 
or are more general in the sense that can constrain different types of particles. 
Below we present a brief summary of our strategy for each search and the results of the validation.

\subsection{ATLAS 1-lepton + 4(1$b$)-jets + $E_T^{\textrm{miss}}$, 21\invfb}

To constrain our scenarios with limits from direct stop production searches we   
simulate the ATLAS 1-lepton + 4(1$b$)-jets + missing energy (MET) search with 20.7\invfb\cite{ATLAS-CONF-2013-037}.
The 95\%~C.L. exclusion bound in the ($m_{\tilde{t}_1}$, $m_{\neutone}$) plane for a SMS of direct stop production with 
$BR(\tone\to t + \neutone)=100\%$ (hereafter called SMS TN) shown in\cite{ATLAS-CONF-2013-037} is comparable to 
the ones obtained with ATLAS all hadronic searches for direct stop  and stop/sbottom production with 
20.5\invfb\ and 20.1\invfb, respectively\cite{ATLAS-CONF-2013-024,ATLAS-CONF-2013-053}. 
It is also comparable to the one given by the CMS 
1-lepton + jets + MET search with 19.5\invfb\cite{CMS-PAS-SUS-13-011}.
The bounds of\cite{ATLAS-CONF-2013-037}  are instead significantly stronger than the ones produced with the ATLAS
2-leptons + jets + MET search with 20.3\invfb\cite{ATLAS-CONF-2013-048}. 
The observed and background yields that we use for our simulation together with the systematic uncertainties are given
in Tables~2--4 of Ref.\cite{ATLAS-CONF-2013-037}. 

\begin{figure}[t]
\centering
\subfloat[]{
\label{fig:a}
\includegraphics[width=0.50\textwidth]{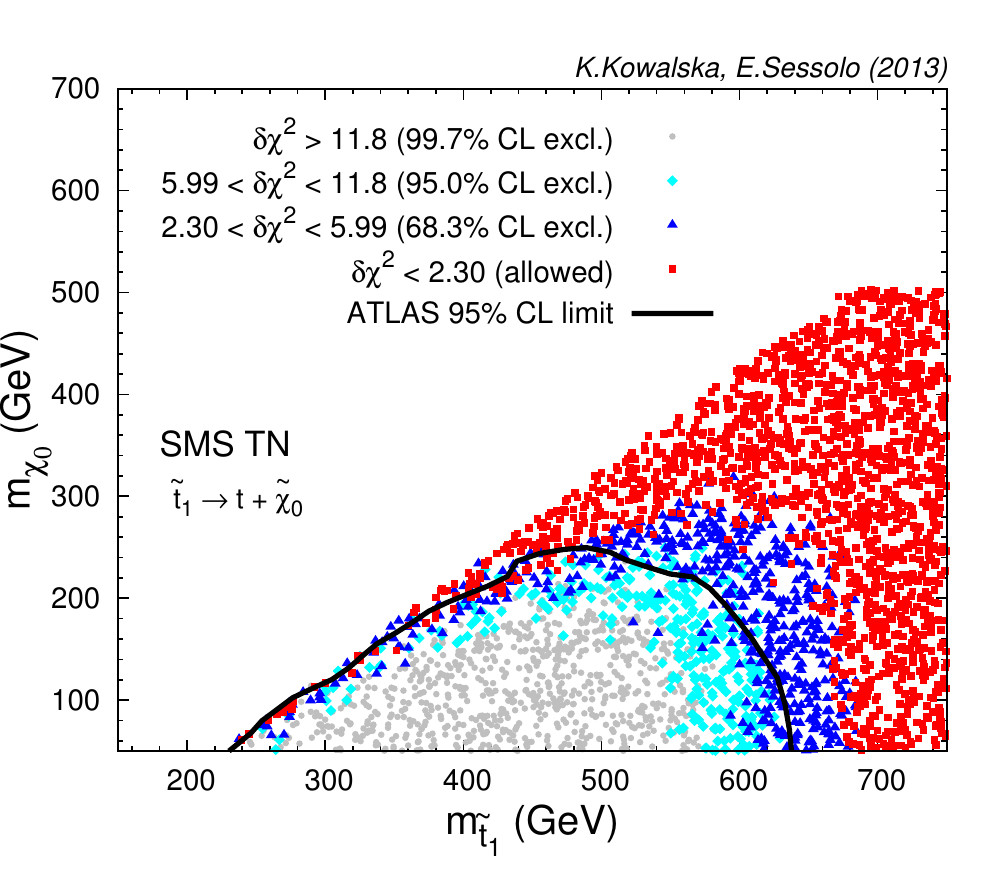}
}
\subfloat[]{
\label{fig:b}
\includegraphics[width=0.50\textwidth]{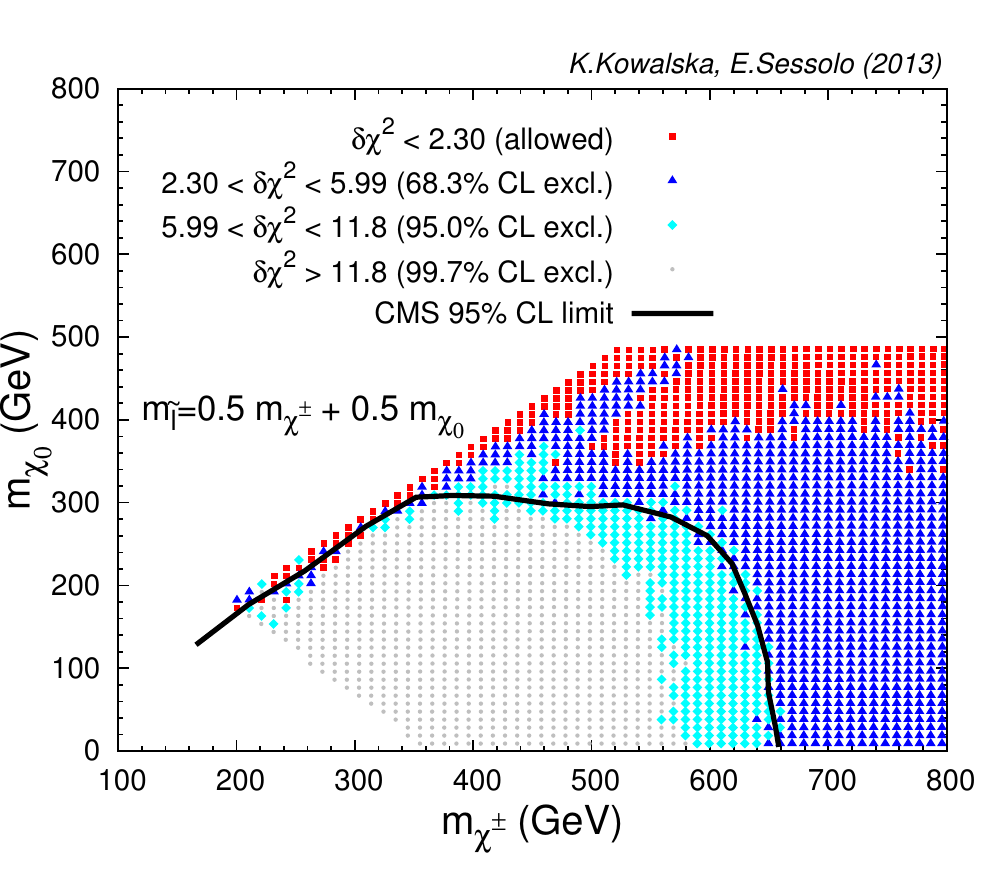}
}
\caption[]{\footnotesize \subref{fig:a} Our simulation of the ATLAS 1-lepton search for direct stop production applied to SMS TN. 
\subref{fig:b} Our simulation of the CMS 3-lepton search for EW production applied to a SMS with $m_{\tilde{l}}=0.5m_{\charone}+0.5m_{\neutone}$. 
Points that are excluded at the 99.7\%~C.L. are showed as gray dots, at the 95.0\%~C.L. as cyan diamonds, 
and at the 68.3\%~C.L. as blue triangles. The points shown as red squares are considered as allowed. 
The solid black lines show the published 95\%~C.L. contours by ATLAS and CMS, which we use for comparison.}
\label{fig:ATLAS_21}
\end{figure}

As a form of validation, we applied our simulation to a sample of 5000
points for which the only light SUSY particles were $\tilde{t}_1$ and a bino-like neutralino.
This was meant to reproduce SMS TN,
for which the ATLAS Collaboration provides a 95\%~C.L. bound in the ($m_{\tilde{t}_1}$, $m_{\neutone}$) plane.
The result of our validation is given in \reffig{fig:ATLAS_21}\subref{fig:a}. 
Gray dots represent the points excluded by our likelihood function at the 99.7\%~C.L., 
cyan diamonds are excluded at the 95.0\%~C.L., 
and blue triangles are excluded at the 68.3\%~C.L. 
The points depicted as red squares are considered as allowed. 
The solid black line shows the 95\%~C.L. ATLAS exclusion limit, which we present for comparison. 

\subsection{CMS 3-leptons + $E_T^{\textrm{miss}}$, 9\invfb}

To constrain our scenarios with limits from direct production of charginos and neutralinos,
we simulate the CMS 3-leptons + MET, EW-production search with 9.2\invfb\cite{CMS-PAS-SUS-12-022}.
Notice that the 95\%~C.L. exclusion bounds published by CMS for the ($m_{\charone}$, $m_{\neutone}$) plane are comparable to the ones obtained 
by the ATLAS 3-leptons + MET search with 20.7\invfb\cite{ATLAS-CONF-2013-035}, given equivalent SMS, 
and are stronger than
the bounds on the same masses obtained by the ATLAS dilepton search with 20.3\invfb\cite{ATLAS-CONF-2013-049}.

The details of our simulation are given in\cite{Fowlie:2013oua}. We repeat that we here updated 
the cross section to the NLO+NLL to increase the accuracy of our calculation.
We limit ourselves to final states with an $ee$ or $\mu\mu$ opposite-sign pair where
the third lepton is either an electron or a muon, which is the box giving the strongest constraints.
The observed and background yields and the systematic uncertainties
are given in Table~1 of\cite{CMS-PAS-SUS-12-022}. 
To validate our likelihood function, we generated a sample of 2500 points 
where the only light particles in the spectrum were 
wino-like \charone\ and \neuttwo, a bino-like \neutone, and 
unified sleptons with mass 
$m_{\slep}=0.5m_{\charone}+0.5m_{\neutone}$. This was meant to reproduce 
one of the SMS for which CMS provided a 95\%~C.L. exclusion bound in the 
($m_{\charone}$, $m_{\neutone}$) plane. The exclusion plot for this 
SMS is presented in \reffig{fig:ATLAS_21}\subref{fig:b}. 
The color code is the same as in \reffig{fig:ATLAS_21}\subref{fig:a}. 
The black solid line represents the 95\%~C.L. exclusion limit by CMS, which we show for comparison.

\subsection{CMS 0-leptons + ($b$-)jets + $E_T^{\textrm{miss}}$ with \alphaT, 12\invfb\label{sec:alphat}}

We implement the bounds on direct production of gluinos, sbottoms and stops with 0 leptons in the final state
by simulating the CMS \alphaT\ search with 11.7\invfb\cite{Chatrchyan:2013lya}. 

The search employs a set of 8 different boxes, with hard jets and MET in the
final states,
and different combinations of $b$-tagged jets. 
It is therefore sensitive to events with different topologies.
For the purpose of this paper we are interested in stop/sbottom production, and 
production of gluinos decaying to squarks of the third generation.
The boxes, together with the number of the observed and background
events provided by the CMS Collaboration, are given in\cite{alphatsite}.

We use this search because of its versatility, and still 
the bounds obtained in the framework of different SMS are among the most constraining 
in the literature. In particular, for gluinos decaying to stops,
the bounds are comparable to the ones
from the CMS $HT$, $b$-jets and MET search with 19.4\invfb\cite{Chatrchyan:2013wxa} and, 
for $m_{\neutone}\lesssim400\gev$, to the bounds from the  
opposite-sign leptons + $b$-jets searches at CMS and ATLAS\cite{Chatrchyan:2012paa,ATLAS-CONF-2013-007} 
and the 3-lepton + $b$-jets search with 19.5\invfb\ at CMS\cite{CMS-PAS-SUS-13-008}. 
However, in this topology, the searches of Refs.\cite{Chatrchyan:2012paa,ATLAS-CONF-2013-007,CMS-PAS-SUS-13-008} 
are more constraining than the 
\alphaT\ search in the  $400\gev\lesssim m_{\neutone}\lesssim 600\gev$ range.
The CMS single-lepton + ($b$-)jets search with 19.4\invfb\cite{CMS-PAS-SUS-13-007}
and the ATLAS 0-lepton + jets + MET search with 20.3\invfb\cite{ATLAS-CONF-2013-047} 
are instead more constraining than the \alphaT\ search by about 200\gev\ for a small neutralino mass.
For gluinos decaying to sbottoms the bounds from the \alphaT\ search are 
the strongest in the literature, comparable to the ones from the CMS $HT$, $b$-jets and MET search. 
For direct sbottom production, the bounds are among the strongest and comparable 
to the bounds from the ATLAS 0-lepton + 2 $b$-jets + MET search with 20.1\invfb\cite{ATLAS-CONF-2013-053}. 
     
Our implementation of the \alphaT\ search is described in detail in\cite{Kowalska:2013hha,Fowlie:2013oua},
with the difference that we here updated the cross section to the NLO+NLL.

\begin{figure}[t]
\centering
\subfloat[]{
\label{fig:a}
\includegraphics[width=0.50\textwidth]{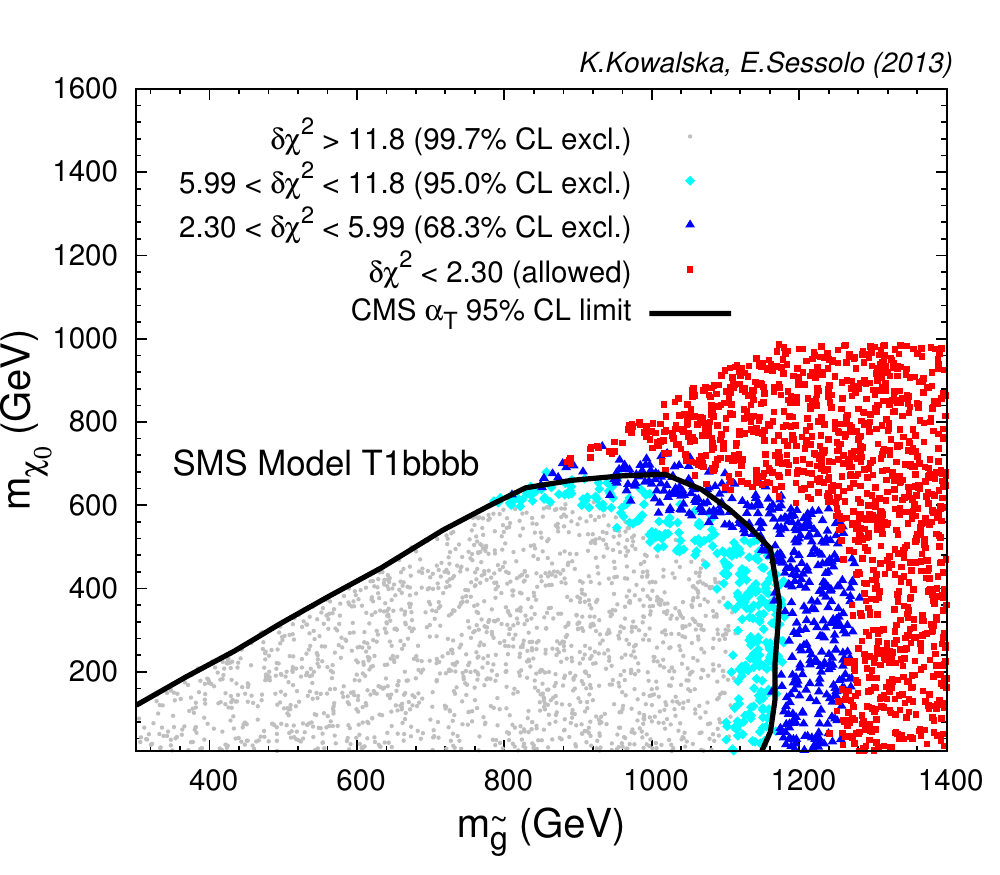}
}
\subfloat[]{
\label{fig:b}
\includegraphics[width=0.50\textwidth]{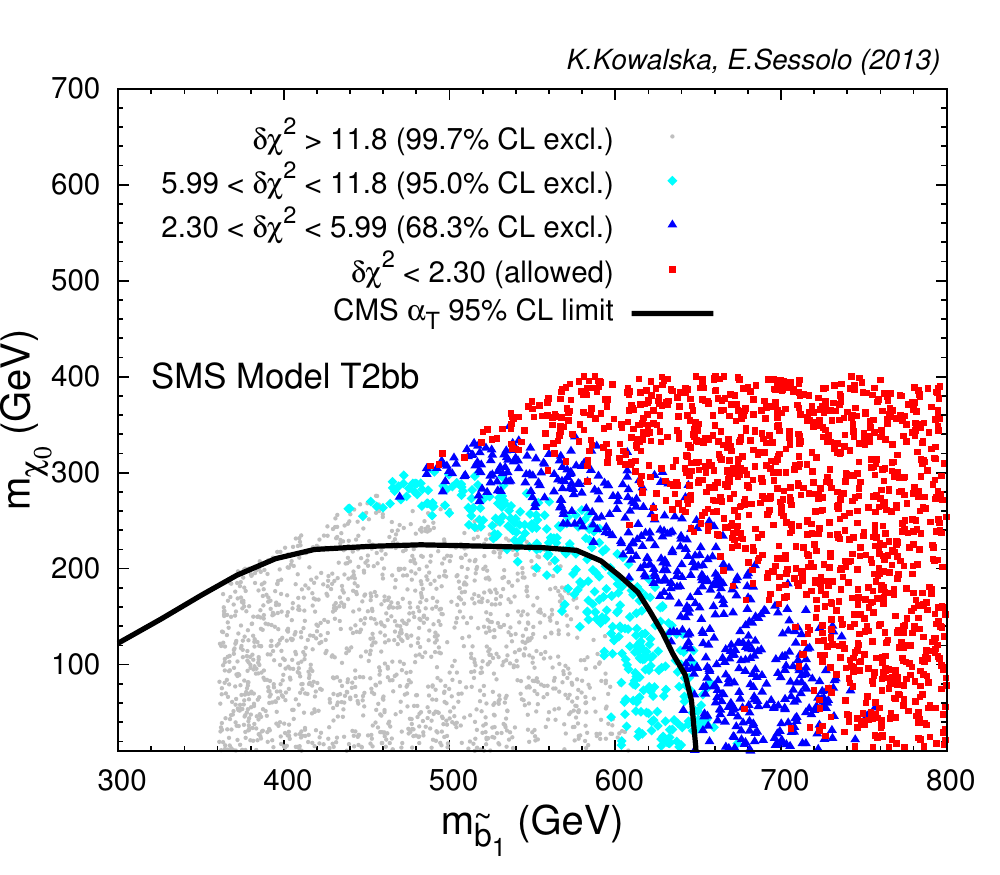}
}
\caption[]{\footnotesize Our simulation of the CMS \alphaT\ search in \subref{fig:a} SMS T1bbbb, and \subref{fig:b} SMS T2bb. The color code is the same as in \reffig{fig:ATLAS_21}.  
The solid black lines show the published 95\%~C.L. contours by CMS, which we use for comparison.}
\label{fig:CMS_12}
\end{figure}

We validated our simulation for direct gluino production on a sample of 5000
points whose spectra presented gluinos, $\tilde{b}_1$, and bino-like neutralinos 
as the sole light particles. This was meant to mimic SMS T1bbbb
for which the CMS Collaboration provided a 95\%~C.L. exclusion bound in the ($m_{\tilde{g}}$, $m_{\neutone}$) plane.
The result of our calculation, compared to the CMS bound, is shown
in \reffig{fig:CMS_12}\subref{fig:a}.

For direct sbottom production we applied the simulation to a sample of points with only
light $\tilde{b}_1$ and bino-like neutralinos, in order to mimic SMS T2bb.
The result, in the ($m_{\tilde{b}_1}$, $m_{\neutone}$) plane, is shown in \reffig{fig:CMS_12}\subref{fig:b}. The color code is the same as in \reffig{fig:ATLAS_21}.
One can see that our likelihood does not reproduce the CMS bound to the desired accuracy in the region with
$m_{\neutone}>200\gev$. We thus remind the reader that our methodology gives only a good approximation
and is not meant to replace the official bounds, which are calculated much more precisely by the experimental collaborations.

\section{Results}\label{res:sec}

In this section we show the impact of the three LHC SUSY searches 
on the parameter space of our scenarios. 
Our conclusions will always be drawn with respect to the 95\%~C.L. bounds obtained from the likelihood function. However, 
as mentioned at the end of \refsec{sec:alphat}, our procedure is an approximation subject to some uncertainty.
We show in our plots the 68.3\%~C.L. and 99.7\%~C.L., which can be loosely interpreted as an estimate of the uncertainty associated with our calculation.

We also calculate in this section the level of fine-tuning for each scenario and discuss the implications of the LHC bounds on some phenomenological observables: 
the Higgs mass, $m_h\simeq125\gev$, Higgs signal rates, the relic density, 
\brbsmumu, \brbxsgamma, and \sigsip.

\subsection{Scenario 1}

As discussed in \refsec{nat:sec}, Scenario~1 is the one characterized by
the smallest number of light SUSY particles. The spectra include light 
$\tilde{t}_{1,2}$, $\tilde{b}_1$, and Higgsino-like, almost degenerate 
\neutone, \neuttwo, and \charone. 

Obviously, the three searches we selected
have different constraining power on the produced spectra.
The ATLAS 1-lepton search is sensitive to stop and sbottom pair production.
The gluinos are too heavy in this scenario, $m_{\tilde{g}}>1730\gev$, to be produced in significant numbers.
The charginos and neutralinos, on the other hand, are degenerate   
so that production of top quarks via processes like $\charone\rightarrow W^{\pm}\neutone$ or 
$\neuttwo\rightarrow Z\neutone$ is highly suppressed.

The limits on $\tilde{t}_1$ are mainly obtained through the $\tilde{t}_1\rightarrow t\neutone$ chain, 
which gives the largest efficiency, 
and the exclusion plot in the ($m_{\tilde{t}_1}$, $m_{\neutone}$) plane looks very similar to \reffig{fig:ATLAS_21}\subref{fig:a},
with only a slightly increased presence of excluded points above the limit obtained in SMS TN. 
This is due to the presence of light sbottoms, which can decay through $\tilde{b}_1\rightarrow t\chi_1^{-}$, 
where the chargino is invisible since it decays softly to the lightest neutralino.

At this point it is worth analyzing the 
possibility of long-lived charginos (in light of the consideration that the lightest neutralinos and chargino are almost degenerate), 
which could provide an alternative and measurable detector signature in the form of long highly ionizing tracks or disappearing charged tracks.
However, we find that this is not an issue in the scenario considered here. 
In fact, in order to make the chargino semistable mass splitting $\Delta m_{\tilde{\chi}^1}\equiv m_{\charone}-m_{\neutone}\lesssim300\mev$ is required\cite{Gunion:1999jr}. 
Such a small mass difference is very difficult to obtain in the case of Higgsino-like LSP, since an additional mass splitting is introduced through radiative corrections, 
unless the gaugino mass parameters are pushed to the multi-TeV regime\cite{Gunion:1999jr}. 
We find that all points in our sample show $\Delta m_{\tilde{\chi}^1}\sim 600\mev-3\gev$.

\begin{figure}[t]
\centering
\subfloat[]{
\label{fig:a}
\includegraphics[width=0.50\textwidth]{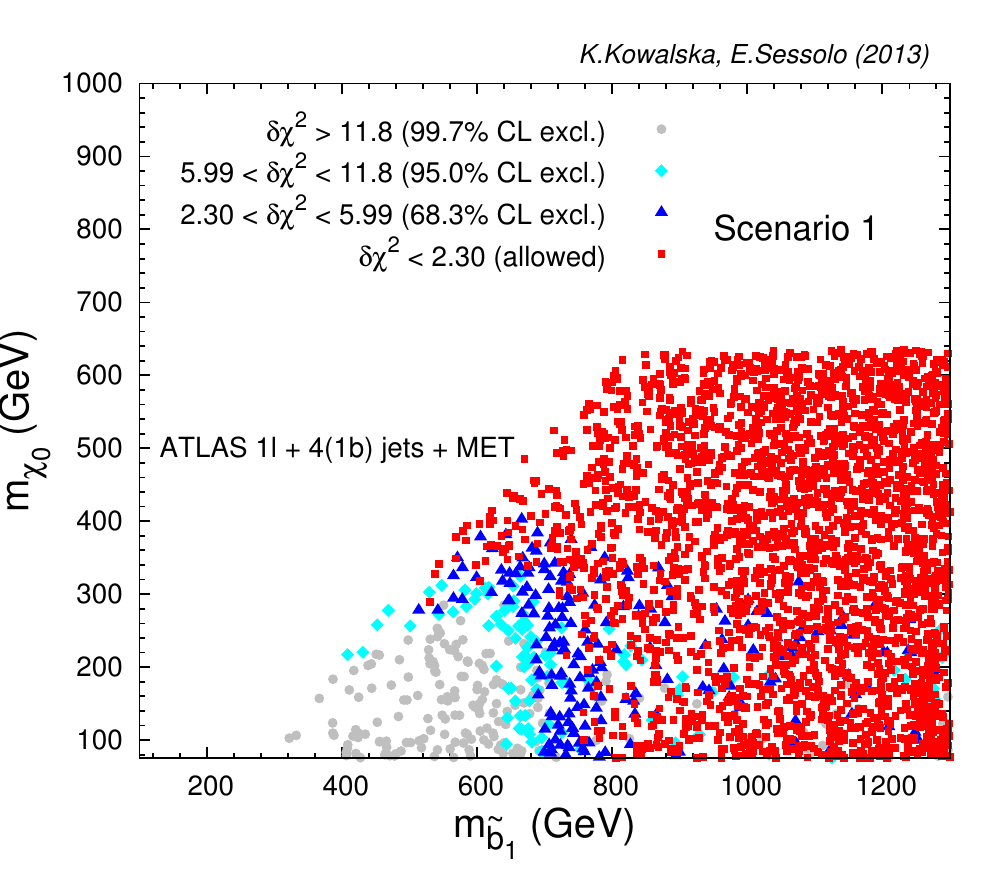}
}\\
\subfloat[]{
\label{fig:b}
\includegraphics[width=0.50\textwidth]{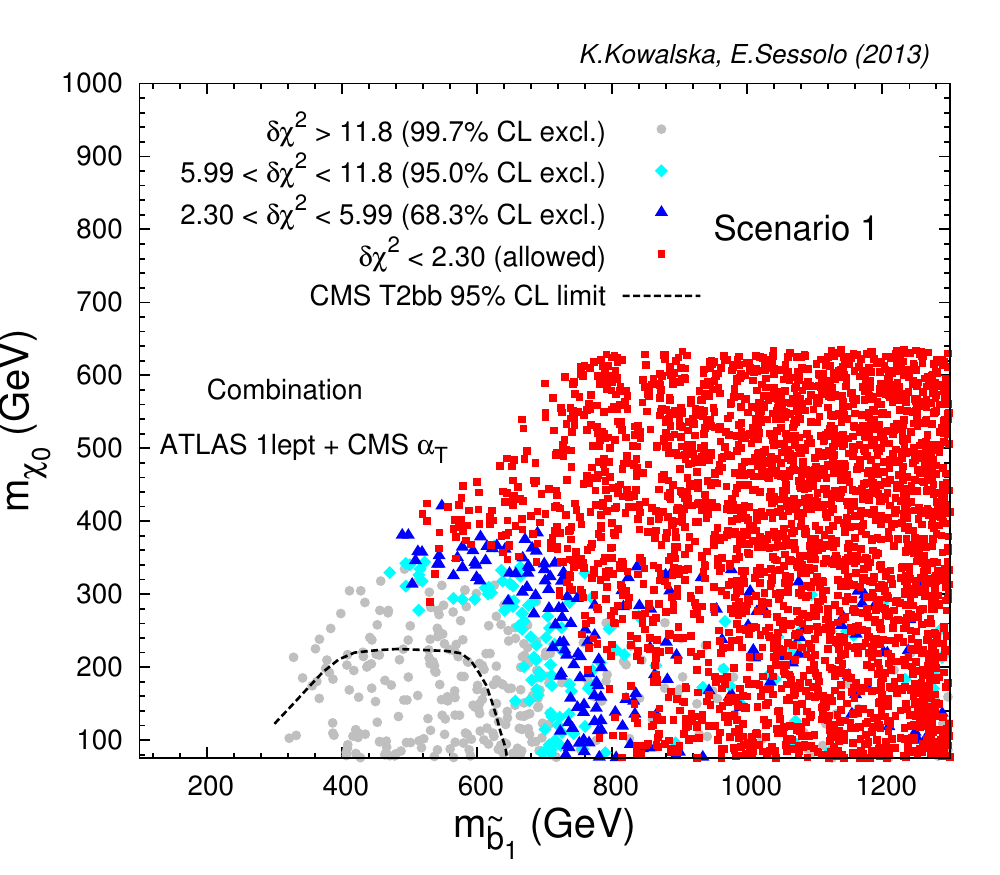}
}
\subfloat[]{
\label{fig:c}
\includegraphics[width=0.50\textwidth]{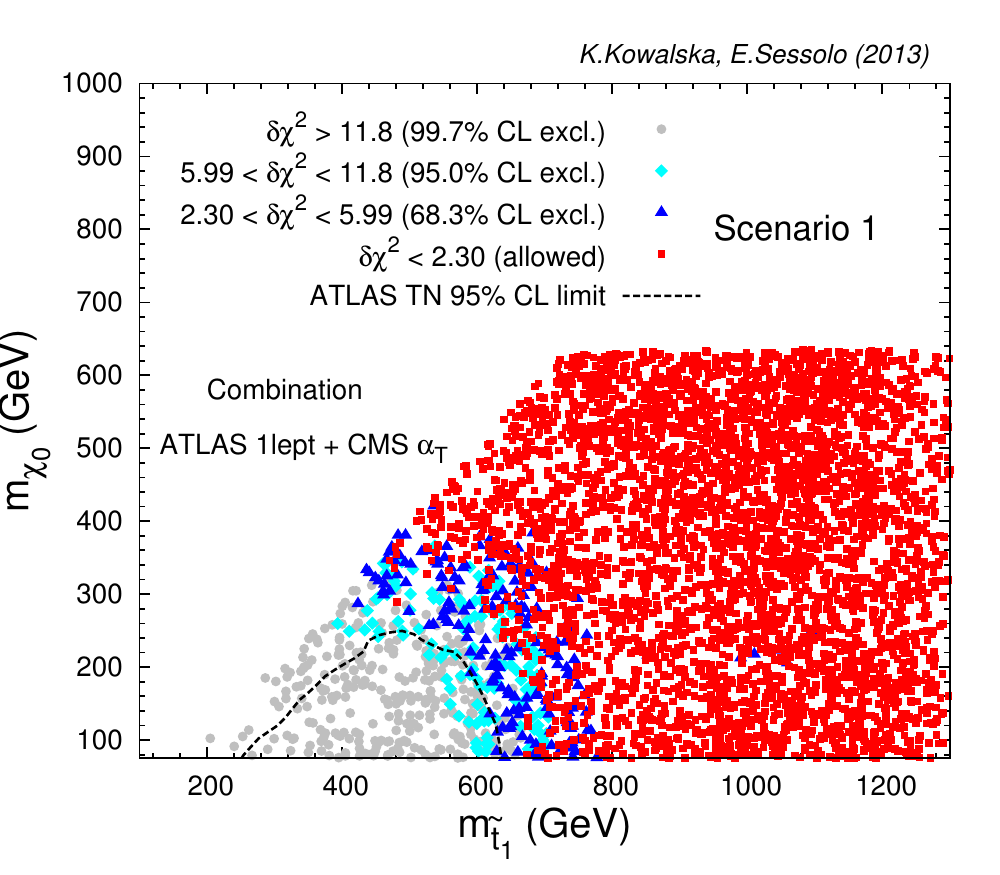}
}
\caption[]{\footnotesize \subref{fig:a} Exclusion levels in the ($m_{\tilde{b}_1}$, $m_{\neutone}$) plane from our simulation of the 
ATLAS 1-lepton search in Scenario~1. \subref{fig:b} Exclusion levels in the ($m_{\tilde{b}_1}$, $m_{\neutone}$) 
plane from our combination of the ATLAS 1-lepton and CMS \alphaT\ searches. The dashed black line shows the published CMS 
\alphaT\ 95\%~C.L. bound in SMS T2bb. \subref{fig:c} Exclusion levels in the ($m_{\tilde{t}_1}$, $m_{\neutone}$) 
plane from our combination of the ATLAS 1-lepton and CMS \alphaT\ searches. The dashed black line shows the published ATLAS 95\%~C.L. 
bound in SMS TN. The color code is the same as in \reffig{fig:ATLAS_21}.}
\label{fig:LHC_S1}
\end{figure}

It is then interesting to notice that in Scenario~1 the ATLAS 1-lepton search can place 
a strong 95\%~C.L. exclusion bound on the mass of the lightest sbottom,
which can be inferred in the ($m_{\tilde{b}_1}$, $m_{\neutone}$) plane from the boundary region between the cyan diamonds and blue triangles 
in \reffig{fig:LHC_S1}\subref{fig:a}.
The light sbottoms are excluded in two different ways: either directly, 
via the $\tilde{b}_1\rightarrow t\chi_1^{-}$ decay chain, as mentioned above, or through the exclusion of stops,
which in this scenario are lighter than the sbottoms.

For final states \textit{without} an isolated lepton with $p_T>25\gev$ (which was instead required by the ATLAS search\cite{ATLAS-CONF-2013-037}), 
the CMS \alphaT\ search can place strong
bounds on the mass of the stops and sbottoms. 
We want to point out here that, while our simulation of 
the ATLAS 1-lepton search
does not provide a neat exclusion limit in the region $m_{\tilde{t}_1}-m_{\neutone}<m_t$,
the \alphaT\ search simulation does. It is known, on the other hand, that this region
is very sensitive to signals from initial state radiation, so that the experimental collaborations generally avoid presenting their limits in that part of the parameter space.  
We have checked that the limits obtained with our \alphaT\ likelihood
in the region $m_{\tilde{t},\tilde{b},\tilde{g}}-m_{\neutone}>100\gev$ are not due 
to spurious initial state jets.
Therefore, while we do not show in this study this region for the ATLAS plots, as it does not give any information,
we will include the parameter space $m_{\tilde{t},\tilde{b},\tilde{g}}-m_{\neutone}>100\gev$ when showing the limits obtained 
with the \alphaT\ search. 

The CMS 3-lepton EW-production search is instead insensitive to this scenario,
since \neutone, \neuttwo, and \charone\ are Higgsino-like, and the resulting spectra 
are highly compressed in the EW sector.

We combine the likelihood functions from the ATLAS 1-lepton and CMS \alphaT\ searches,
which are obviously statistically independent, to derive 95\%~C.L. bounds 
on the lightest stops and sbottoms in Scenario~1. They can be inferred from the boundary between 
the cyan diamonds and blue triangles in \reffig{fig:LHC_S1}\subref{fig:b}
and in \reffig{fig:LHC_S1}\subref{fig:c}, for the ($m_{\tilde{b}_1}$, $m_{\neutone}$)
and ($m_{\tilde{t}_1}$, $m_{\neutone}$) planes, respectively.
For comparison, the dashed black line in \reffig{fig:LHC_S1}\subref{fig:b} 
gives the official 95\%~C.L. in SMS T2bb for the CMS \alphaT\ search, which is one of the SMS we used for validation of our procedure as described in \refsec{sec:alphat}. 
Equivalently, the dashed black line in \reffig{fig:LHC_S1}\subref{fig:c} gives the official 95\%~C.L. in SMS TN for the ATLAS 1-lepton search.

One can see in \reffig{fig:LHC_S1}\subref{fig:b} that, for a neutralino in the mass range 
$75\gev\leq\neutone\lesssim 300\gev$, $m_{\tilde{b}_1}\lesssim 700\gev$ is excluded at the 
95\%~C.L. Figure~\ref{fig:LHC_S1}\subref{fig:c} shows that, for $75\gev\leq\neutone\lesssim 250\gev$, 
$m_{\tilde{t}_1}\lesssim 650\gev$ is excluded at the 95\%~C.L. 

\begin{figure}[t]
\centering
\subfloat[]{
\label{fig:a}
\includegraphics[width=0.50\textwidth]{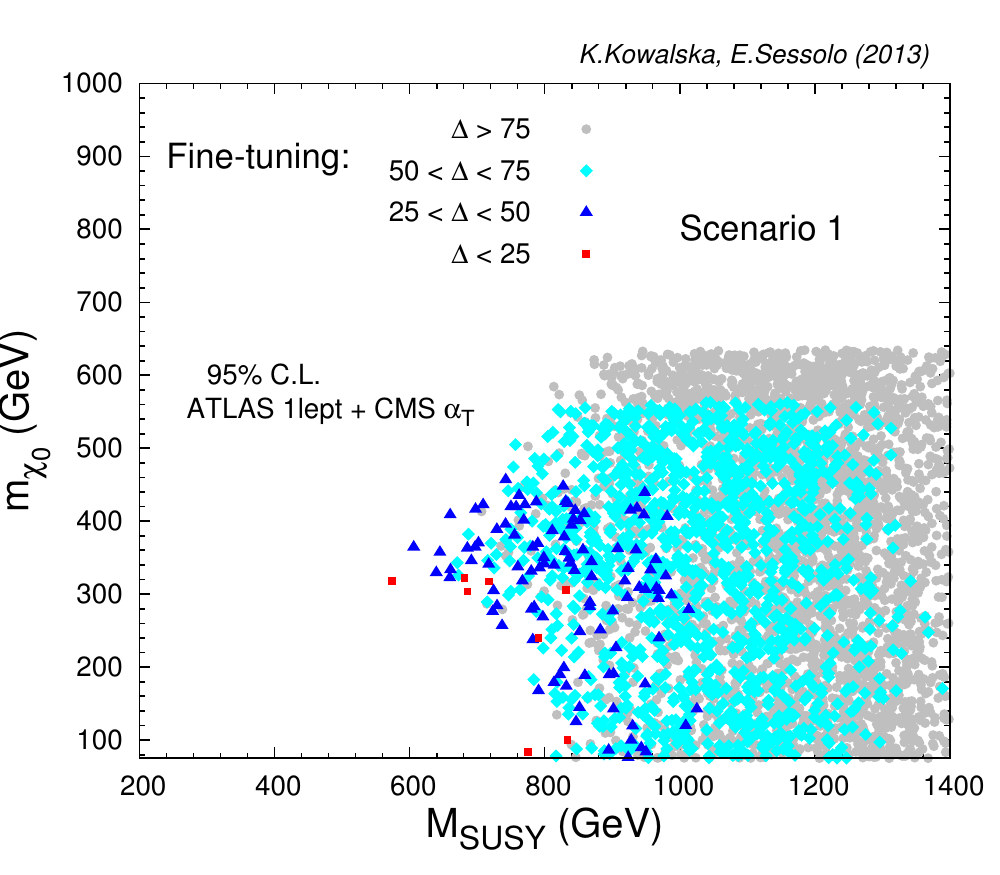}
}
\subfloat[]{
\label{fig:b}
\includegraphics[width=0.50\textwidth]{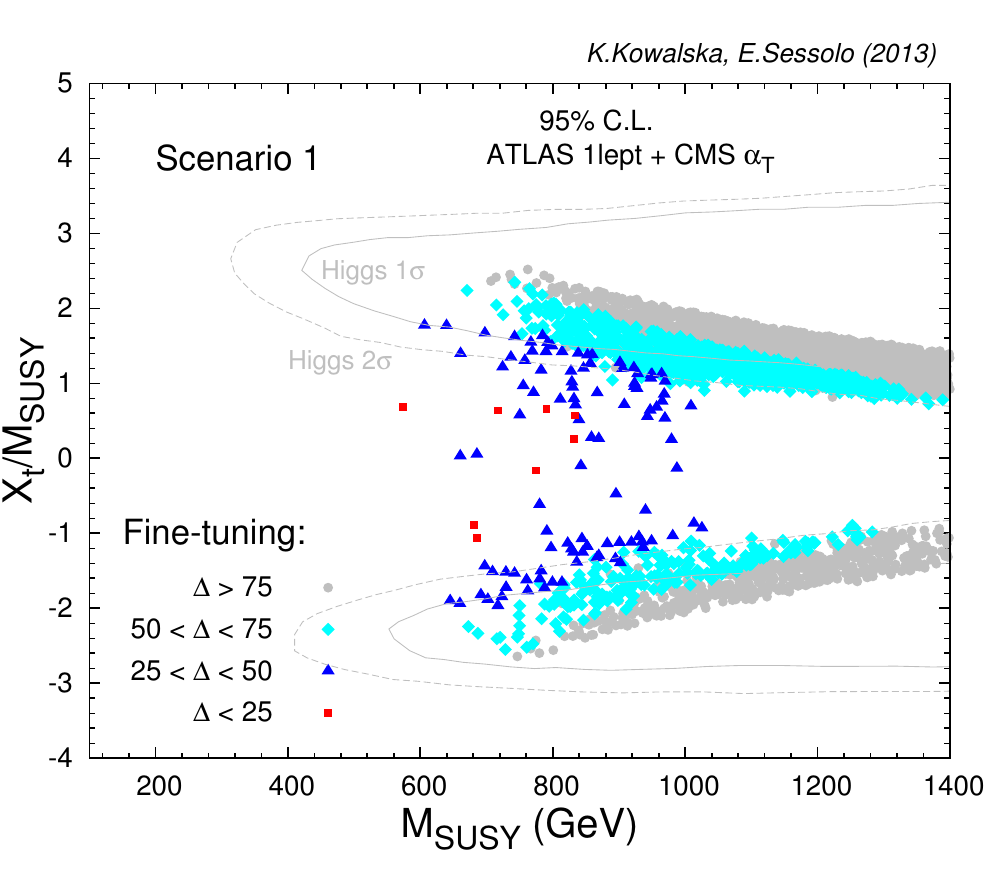}
}
\caption[]{\footnotesize  Scatter plot of the fine-tuning 
measure $\Delta$ for the points that are not 
excluded at the 95\%~C.L. by the LHC for Scenario~1 in \subref{fig:a} the (\msusy, $m_{\neutone}$) plane and \subref{fig:b} the (\msusy, $X_t/\msusy$) plane. Red squares correspond to $\Delta\leq25$, blue triangles to $25<\Delta\leq 50$, cyan diamonds to $50<\Delta\leq 75$, and gray dots to $\Delta>75$. The solid (dashed) gray contours indicate the approximate 
$1\sigma$ ($2\sigma$) window for the Higgs mass.}
\label{fig:FT_S1}  
\end{figure}

Note that the results presented in \reffig{fig:LHC_S1}\subref{fig:b} and \reffig{fig:LHC_S1}\subref{fig:c} 
are in a good agreement with Fig.~7 of Ref.\cite{Kribs:2013lua}, where the limits from five CMS and ATLAS stop/sbottom 
searches were combined for a model with light and almost degenerate $\tilde{t}_{1}$, $\tilde{t}_{2}$ and $\tilde{b}_{1}$ in the spectrum. 
A slightly weaker bound on $\tilde{t}_{1}$ comes in our case from the fact that here stops and sbottoms are not degenerate, 
and the sbottom is in most cases heavier than the lightest stop. 
This mass hierarchy also explains the presence of points excluded at 95\% C.L. for $m_{\tilde{b}_1}>1\tev$ in \reffig{fig:LHC_S1}\subref{fig:b}, 
which are characterized by $\tilde{t}_{1}$ light enough to be tested by the LHC.

We then calculate $\Delta$ according to Eq.~(\ref{finetune}), for a conservative value $\Lambda=10\tev$. 
The result is shown in \reffig{fig:FT_S1}\subref{fig:a} in the (\msusy, $m_{\neutone}$) plane ($\msusy=\sqrt{m_{\tilde{t}_1}m_{\tilde{t}_2}}$) 
for the points that are not excluded 
at the 95\%~C.L. by the LHC. 
One can see a handful of points characterized by $\Delta\leq 25$ and some more 
with $25<\Delta\leq 50$. We could not find any points with $\Delta\leq20$, as they are all disfavored by the LHC.

The features of the points with the lowest fine-tuning can be inferred by comparing
\reffig{fig:FT_S1}\subref{fig:a} with \reffig{fig:FT_S1}\subref{fig:b}, where we show the fine-tuning distribution in the 
(\msusy, $X_t/\msusy$) plane, with $X_t=A_t-\mu\cot\beta$. We also plot 
in \reffig{fig:FT_S1}\subref{fig:b} the approximate $1\sigma$ (solid contour) and $2\sigma$ (dashed contour) windows for the Higgs mass.
Note that the points at $\msusy\lesssim1000\gev$ and with the smallest stop mixing are the points that do not belong to the $2\sigma$ window for the Higgs mass, as explained at the end of
\refsec{nat:sec}. 

The points with $\Delta\leq 25$ are characterized 
by $\msusy\lesssim 850\gev$, $m_{\neutone}\approx\mu\lesssim 320\gev$, and small stop mixing, $|A_t|\lesssim1000\gev$.
It is therefore safe to say that these points are likely to be excluded
in the early stages of the LHC $\sqrt{s}=14\tev$ run. 
Figure~\ref{fig:FT_S1}\subref{fig:a} also shows many points with $\Delta>25$ in the same region of the (\msusy, $m_{\neutone}$) plane. Larger fine-tuning is for those points due to increasing stop mixing, as can be inferred from \reffig{fig:FT_S1}\subref{fig:b}.

\bigskip

\begin{figure}[t]
\centering
\subfloat[]{
\label{fig:a}
\includegraphics[width=0.50\textwidth]{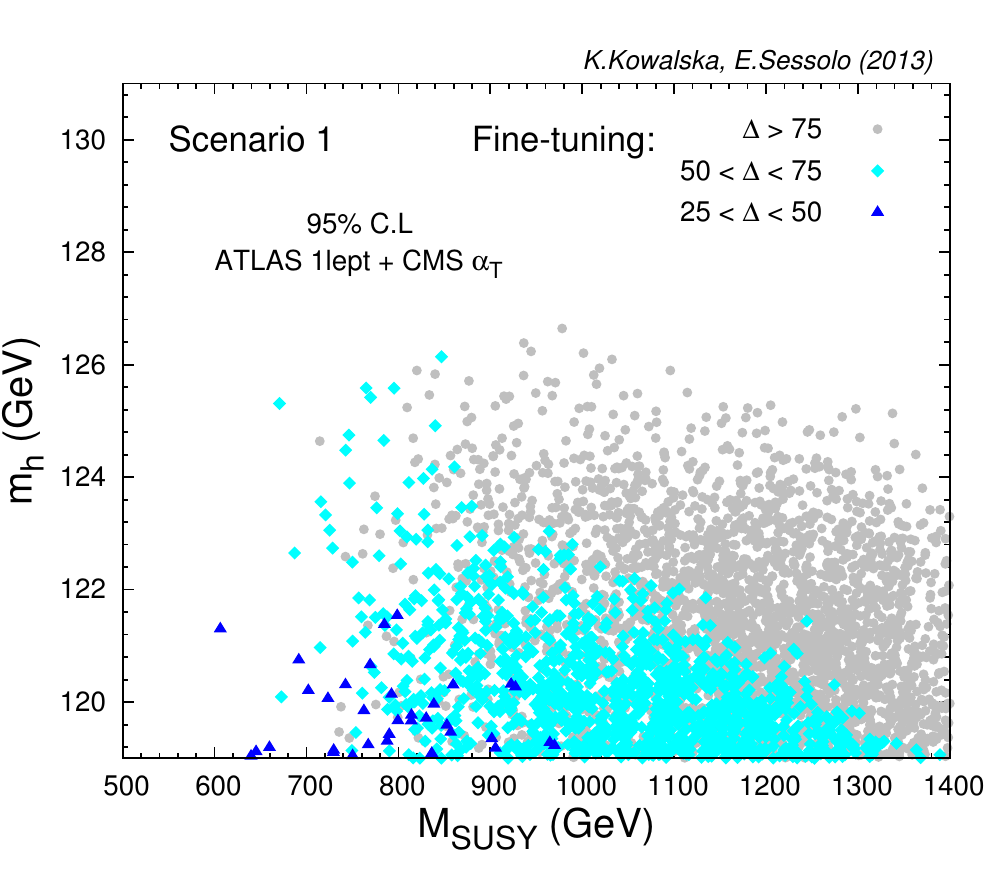}
}
\subfloat[]{
\label{fig:b}
\includegraphics[width=0.50\textwidth]{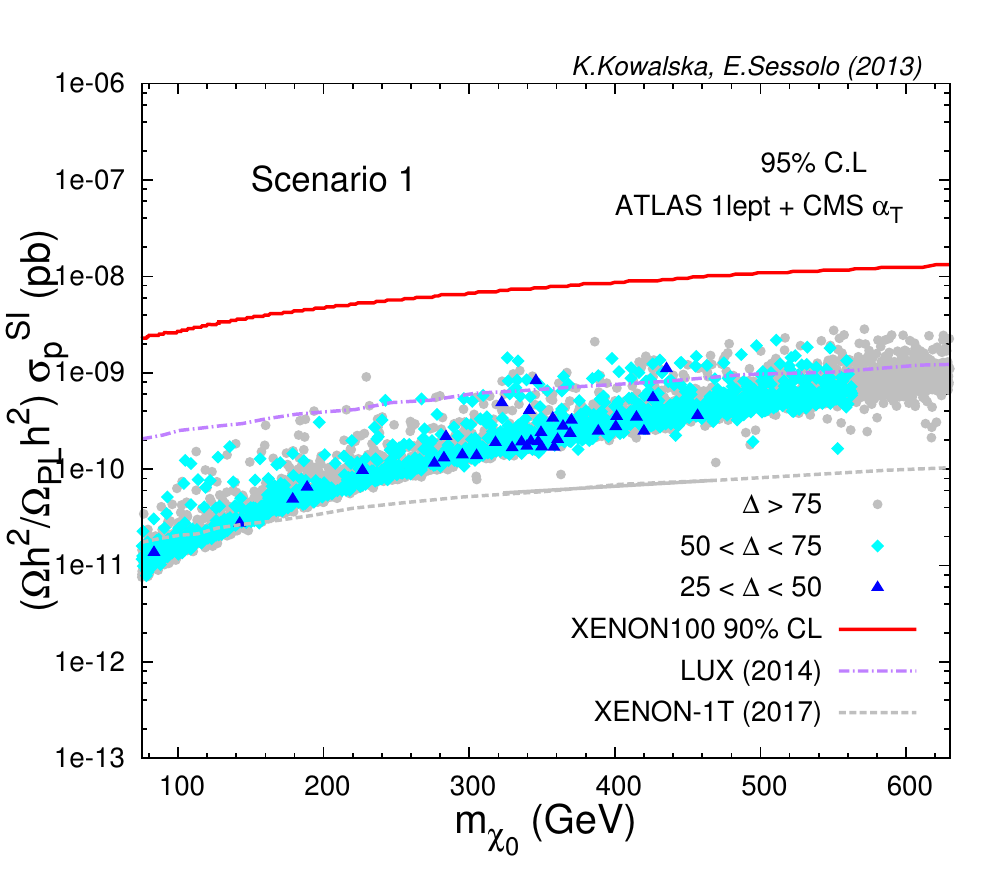}
}
\caption[]{\footnotesize Scatter plot of the fine-tuning measure $\Delta$ for the points that are not 
excluded at the 95\%~C.L by the LHC and are characterized by $\mhl\geq119\gev$ ($2\sigma$ window) in Scenario~1 in \subref{fig:a} 
the (\msusy, $m_h$) plane and \subref{fig:b} the ($m_{\neutone}$, $\Omega_{\chi} h^2/\Omega_{\textrm{Planck}}h^2\cdot\sigsip$) plane. 
The solid red line shows the 90\%~C.L. bound from XENON100, while the dot-dashed purple and dashed gray lines 
show future sensitivities at LUX and XENON1T, respectively. The color code is the same as in \reffig{fig:FT_S1}.}
\label{fig:constr_S1}
\end{figure}

In \reffig{fig:constr_S1}\subref{fig:a}, we show a scatter plot in the (\msusy, $m_h$) plane of the fine-tuning measure 
$\Delta$ for the points allowed by the LHC constraints.
We plot only the points that belong to the $2\sigma$ window for the Higgs mass, with theoretical and experimental uncertainties added in quadrature,
as explained at the end of \refsec{nat:sec}. 

As was anticipated in \reffig{fig:FT_S1}\subref{fig:b}, \reffig{fig:constr_S1}\subref{fig:a} shows that none of the points with the lowest fine tuning (red squares in \reffig{fig:FT_S1}\subref{fig:b}) have \mhl\ consistent 
with the experimental value within $2\sigma$. 
As a matter of fact, those points show low Higgs masses, in the range $\mhl\simeq 110-115\gev$.
In this sense we agree with\cite{Hall:2011aa,CahillRowley:2012rv,CahillRowley:2012kx,Blanke:2013uia,Kribs:2013lua},
i.e., with the possible exclusion of the region with compressed spectra, 
there seems to be no room for points with small $\Delta$
given the present status of LHC searches and the measurement of the Higgs mass.
Moreover, the value of $m_h$ can be accommodated for points with $25<\Delta\leq 50$
only with the help of a considerable theoretical error added to the numerical calculation,
which is performed with \texttt{softsusy} in this study. 

As was also mentioned in \refsec{nat:sec}, by construction all the points that survive 
the LHC and Higgs mass bounds, which are shown in \reffig{fig:constr_S1}\subref{fig:a},
satisfy the constraints on \brbxsgamma\ and \brbsmumu\ at $2\sigma$ (we adopt the central values and uncertainties
given in 
Table~2 of Ref.\cite{Fowlie:2013oua}). Therefore, we refrain in this study from showing distributions for those observables.

When comparing the Higgs signal rates, \rgg\ and \rzz, to their experimentally measured values by ATLAS and CMS\cite{ATLAS-CONF-2013-012,*CMS-PAS-HIG-13-001}, 
we find that the CMS determinations $\rgg=0.77\pm 0.27$ and $\rzz=0.91\pm 0.30$ do not affect the parameter space at all, as 
100\% of the points fall into the $2\sigma$ intervals.
As a matter of fact, only the ATLAS determination in the $\gamma\gamma$ channel, $\rgg=1.65\pm 0.35$,
has some impact on the parameter space of Scenario~1, excluding about 26\% of the points at the $2\sigma$ level. However, 
for those points we do not observe any correlation between the exclusion level and the parameters relevant
for the study of fine-tuning.  
  
The relic density constraint deserves more consideration.
In Scenario~1 the lightest neutralino is Higgsino-like and its mass is 
approximately equal to the value of the $\mu$ parameter.
The relic density is in this case easily expressed in terms of $\mu$,
$\Omega h^2\approx 0.1\cdot(\mu/\textrm{TeV})^2$\cite{ArkaniHamed:2006mb}.
For values in our scanned range, $75\gev<\mu\leq630\gev$, the relic density 
yields for all points a value between 0.001 and 0.05.
One can consider the case where the neutralino is not the sole component of dark matter;
see, e.g.,\cite{Baer:2013vpa}. In this case, assuming that the local density 
of neutralinos is obtained from the total local density by rescaling with a correction factor,
$\Omega_{\chi} h^2/\Omega_{\textrm{Planck}}h^2$, we rescale the value of the SI neutralino-proton
scattering cross section and in this way account for the weakening of the signal at the underground detector.
We show in \reffig{fig:constr_S1}\subref{fig:b} the scatter plot of the fine-tuning measure in the 
($m_{\neutone}$, $\Omega_{\chi} h^2/\Omega_{\textrm{Planck}}h^2\cdot\sigsip$) plane for the points that 
satisfy the Higgs mass and LHC constraints (the points of \reffig{fig:constr_S1}\subref{fig:a}).
As expected, the value of \sigsip\ is independent of the level of fine-tuning and 
the distribution of points agrees with the results of\cite{Baer:2013vpa}, in which the same calculation 
was performed for a natural NUHM2 type of model. 

We compare our scattered points with the 90\%~C.L. bound from XENON100\cite{Aprile:2012nq} (solid red line) 
and we also show sensitivities at LUX\cite{Akerib:2012ys} (dot-dashed purple line) and XENON1T\cite{Aprile:2012zx} (dashed gray line). 
The latter in particular should be able to test a very significant part of the parameter space of the model.\footnote{The theoretical uncertainties on \sigsip\ 
due to the pion-nucleon $\Sigma$ term can significantly reduce the impact of the XENON100 limit, 
as well as the prospects for the future sensitivities, as shown in detail in\cite{Fowlie:2013oua}.}

\subsection{Scenario 2}

\begin{figure}[t]
\centering
\subfloat[]{
\label{fig:a}
\includegraphics[width=0.50\textwidth]{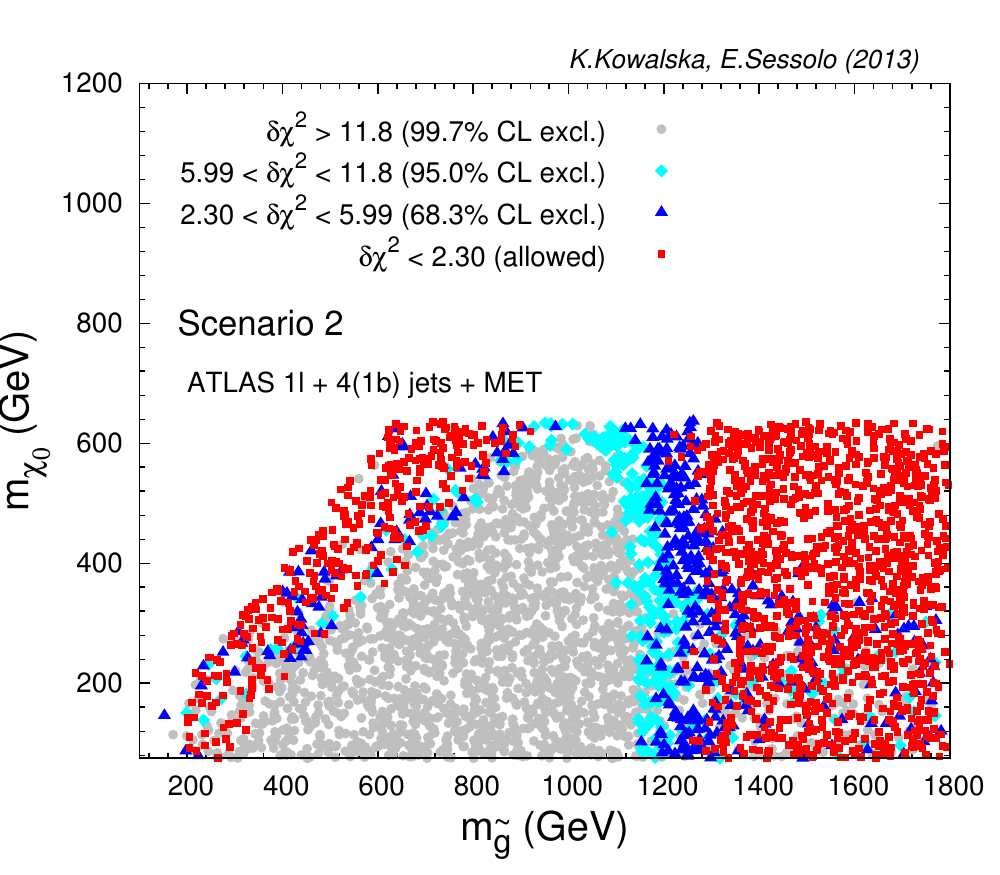}
}
\subfloat[]{
\label{fig:b}
\includegraphics[width=0.50\textwidth]{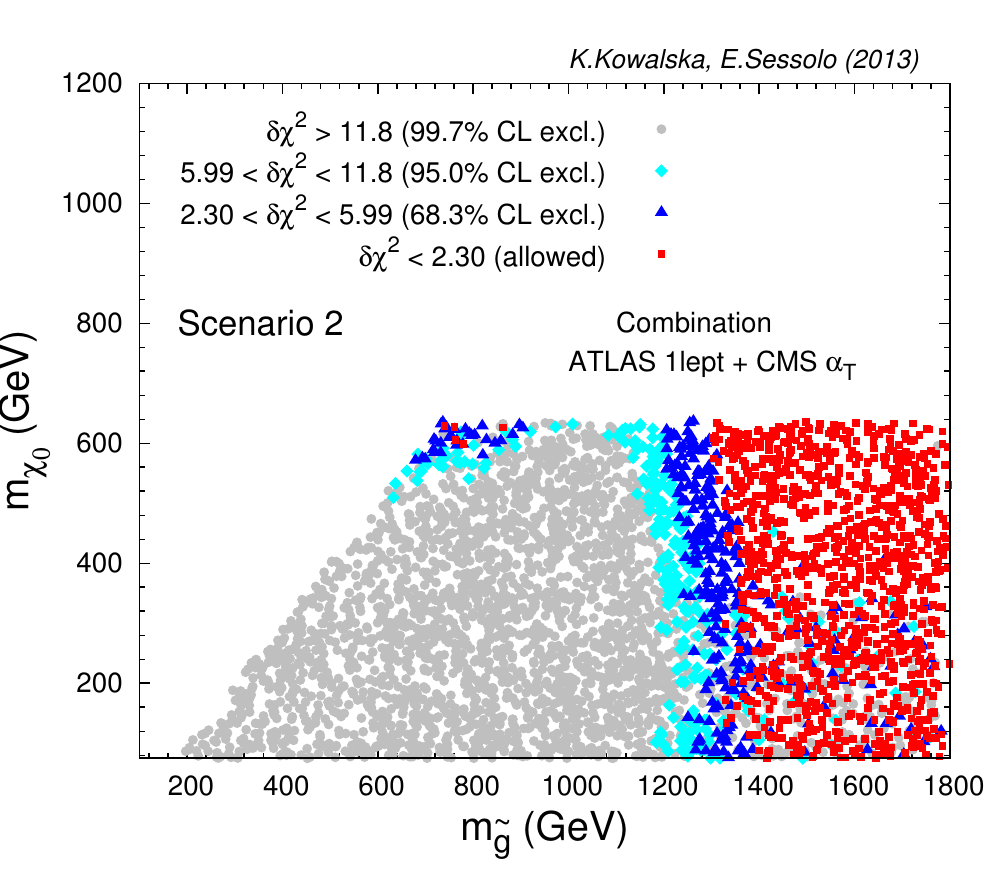}
}\\
\subfloat[]{
\label{fig:c}
\includegraphics[width=0.50\textwidth]{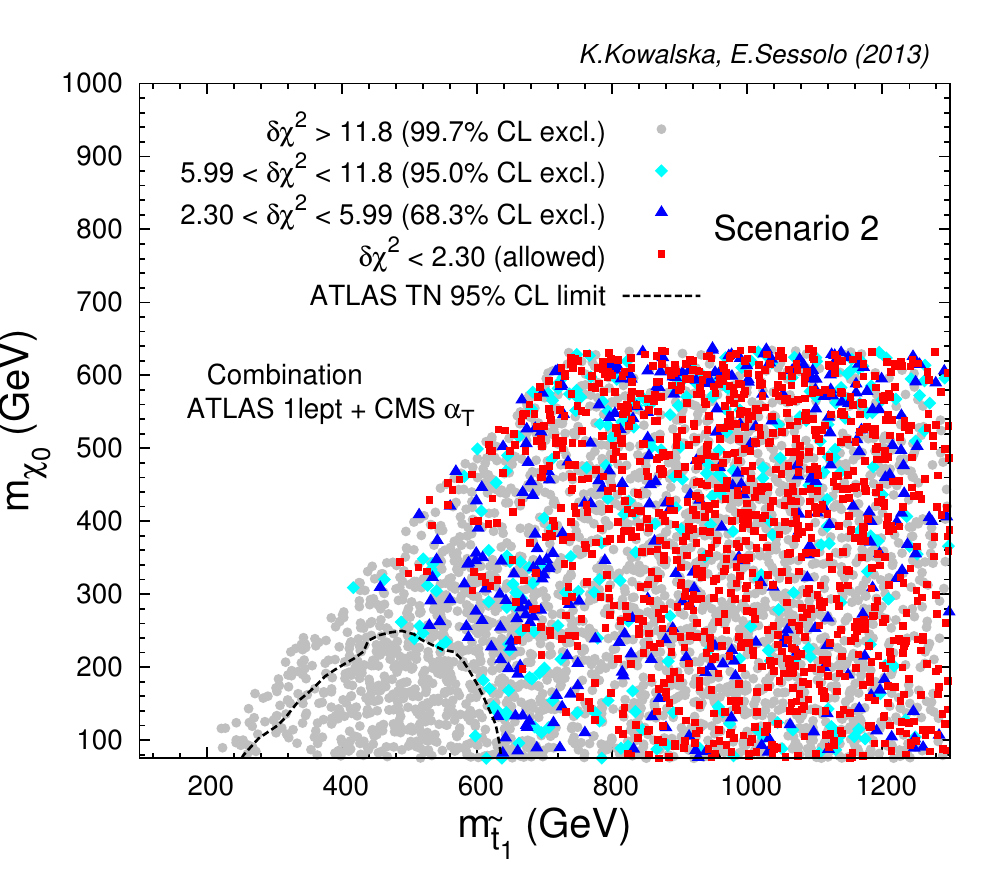}
}
\subfloat[]{
\label{fig:d}
\includegraphics[width=0.50\textwidth]{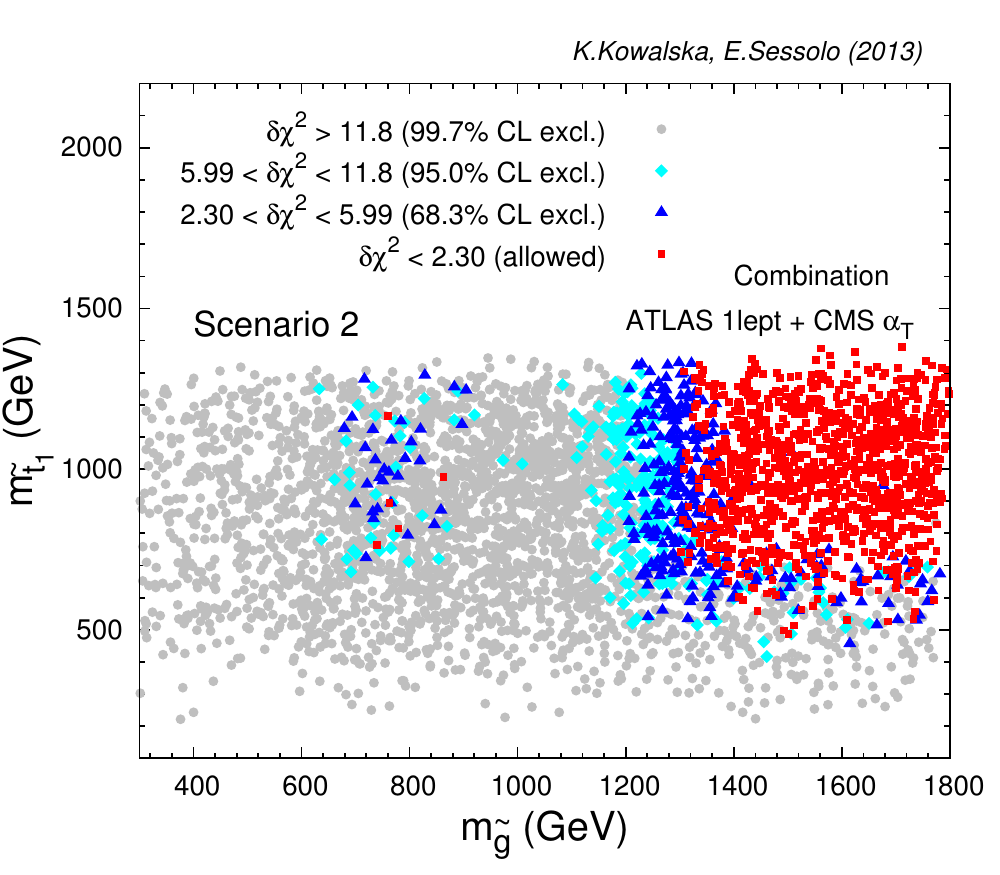}
}
\caption[]{\footnotesize Exclusion levels in the ($m_{\tilde{g}}$, $m_{\neutone}$) plane from \subref{fig:a} 
our simulation of the ATLAS 1-lepton search and \subref{fig:b} our combination of the ATLAS 1-lepton and CMS \alphaT\ 
searches in Scenario~2. Exclusion levels from the same combination in \subref{fig:c} the ($m_{\tilde{t}_1}$, $m_{\neutone}$) 
plane and \subref{fig:d} the ($m_{\tilde{g}}$, $m_{\tilde{t}_1}$) plane. The color code is the same as in \reffig{fig:ATLAS_21}.}
\label{fig:LHC_S3}
\end{figure}

In this scenario the spectra are characterized by the same set of
particles as in Scenario~1, but this time the gluino can be lighter than
the squarks of the third generation and within reach of the LHC. We will
see that this property makes this scenario more constrained than Scenario~1.
On the other hand, we do not expect variations in the overall level of
fine-tuning,
since already in Scenario~1 the contribution to $\Delta$ of the decoupled
gluino was generally less important than the ones due to $\mu$ or the
third generation squarks.

The CMS \alphaT\ search places limits on gluinos decaying to stops and
sbottoms,
as discussed in \refsec{sec:alphat}. On the other hand, the
mass of the gluino is also strongly constrained by the ATLAS 1-lepton search,
in spite of the fact that the latter was designed for detection of
directly produced stops.
We show in \reffig{fig:LHC_S3}\subref{fig:a} the impact of the ATLAS search
on the ($m_{\tilde{g}}$, $m_{\neutone}$) plane, for which ATLAS did not
provide an official
exclusion bound. Neglecting the region on the left of the plot,
for which the spectrum is compressed,
we derive a strong bound, $m_{\tilde{g}}\gsim 1200\gev$, from the ATLAS
1-lepton search in
this scenario.
Interestingly, this limit is in good agreement with the bound obtained in
the same plane by the
CMS single-lepton + $b$-jets + MET search\cite{CMS-PAS-SUS-13-007}.

We can now statistically combine the ATLAS 1-lepton and CMS \alphaT\
searches to provide a stronger bound on the
($m_{\tilde{g}}$, $m_{\neutone}$) plane, which can be inferred
in \reffig{fig:LHC_S3}\subref{fig:b}
from the boundary between the cyan diamonds and blue triangles. 
Although strongly dominated by the constraining power of the \alphaT\ search, 
the exclusion in \reffig{fig:LHC_S3}\subref{fig:b} is
stronger than in each individual case.

In \reffig{fig:LHC_S3}\subref{fig:c}, we show the exclusion plot from our statistical combination in the ($m_{\tilde{t}_1}$, $m_{\neutone}$) 
plane. There are many more points excluded at the 95\%~C.L. than in 
Scenario~1, due to the presence of a light gluino in the spectrum. This makes it more difficult than in Scenario~1 to find allowed points with 
$m_{\tilde{t}_1}\lesssim 650\gev$. 

We summarize the LHC limits for Scenario~2 in \reffig{fig:LHC_S3}\subref{fig:d} where
we show the exclusion plot in the ($m_{\tilde{g}}$, $m_{\tilde{t}_1}$) plane. 
Most points with $m_{\tilde{g}}\leq1200\gev$ are excluded
independently of the value of the stop mass. The points that are not excluded in the region $m_{\tilde{g}}\simeq800\gev$ are the ones close to the compressed spectra region
for the gluinos, shown on the top left in \reffig{fig:LHC_S3}\subref{fig:b}.
The points in the range $400\gev\lesssim m_{\tilde{t}_1}\lesssim 600\gev$
that are not excluded
at the 95\%~C.L. (red squares and blue triangles) are the points close to
the compressed spectra region for the stops,
shown in \reffig{fig:LHC_S3}\subref{fig:c} in the range $300\gev\lesssim
m_{\neutone}\lesssim 400\gev$.

The limits do not change by including the CMS 3-lepton EW-production search as, similarly to Scenario~1, the neutralino
is Higgsino-like and the 3-lepton search is not sensitive to
spectra compressed in the EW sector.

\begin{figure}[t]
\centering
\subfloat[]{
\label{fig:a}
\includegraphics[width=0.50\textwidth]{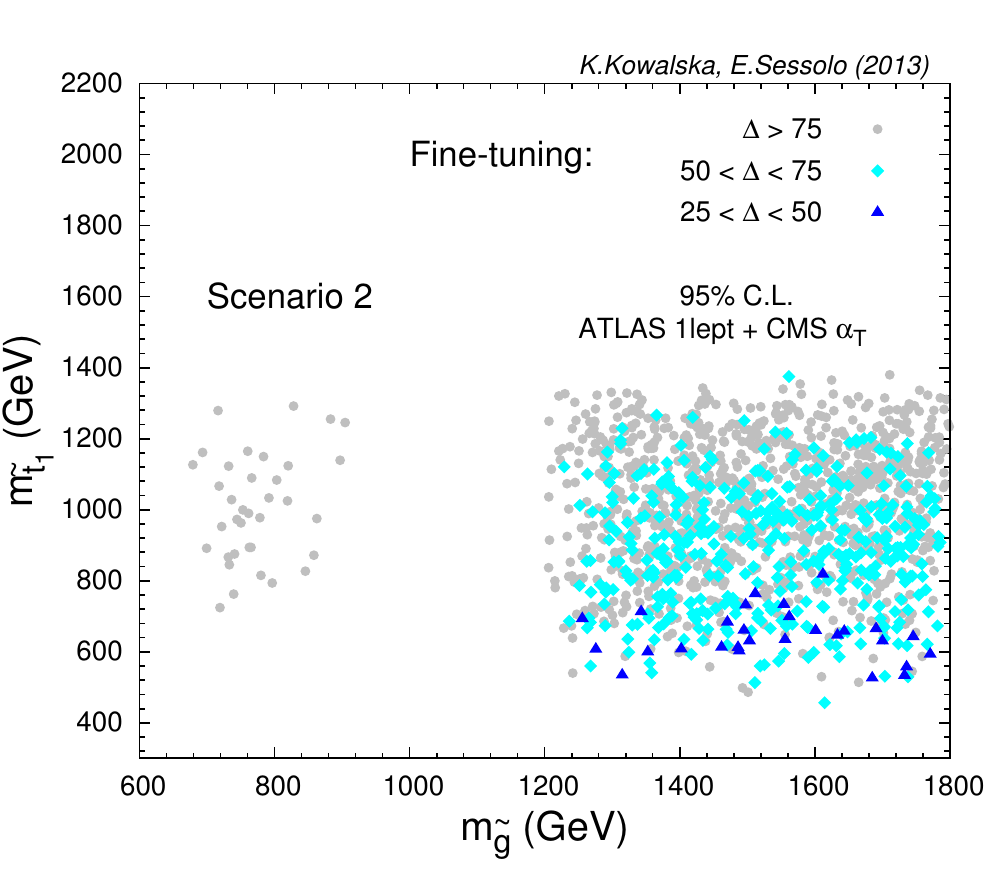}
}
\subfloat[]{
\label{fig:b}
\includegraphics[width=0.50\textwidth]{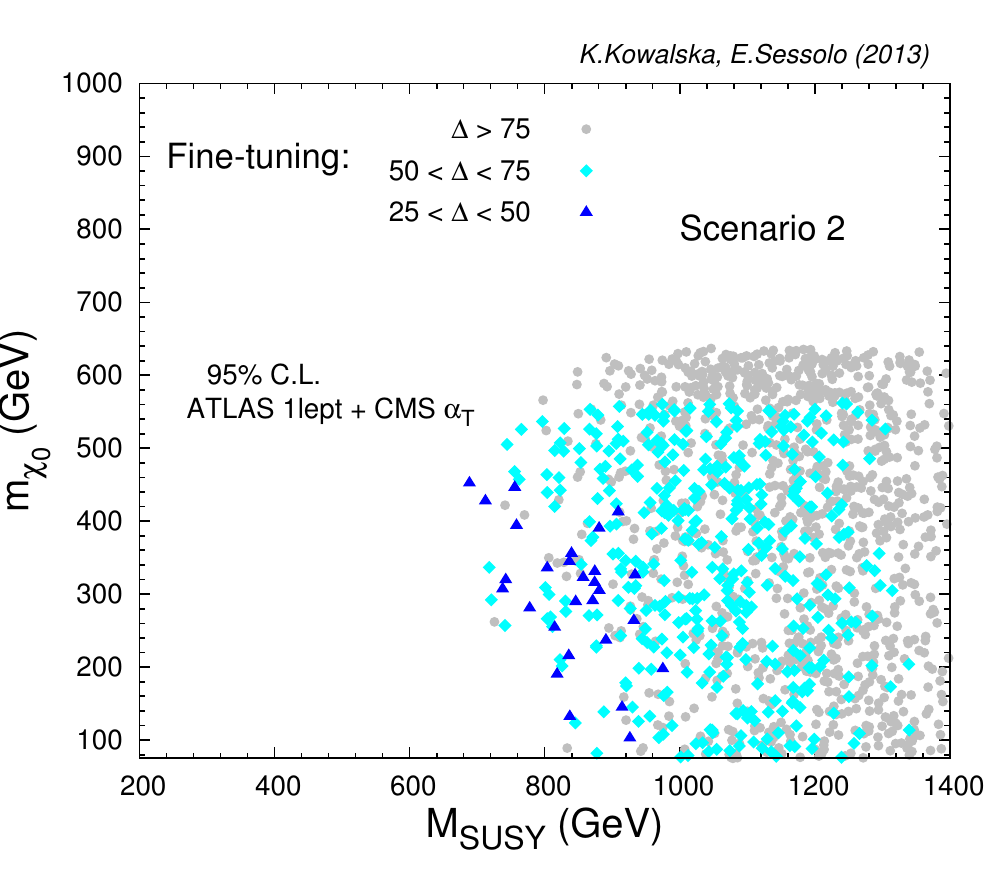}
}
\caption[]{\footnotesize Scatter plot of the fine-tuning measure $\Delta$ for the points that are not excluded at 
the 95\%~C.L. by the LHC in Scenario~2 for \subref{fig:a} the ($m_{\tilde{g}}$, $m_{\tilde{t}_1}$) plane and 
\subref{fig:b} the (\msusy, $m_{\neutone}$) plane. The color code is the same as in \reffig{fig:FT_S1}.}
\label{fig:FT_S3}
\end{figure}

In \reffig{fig:FT_S3}\subref{fig:a} we show the distribution of the
fine-tuning measure $\Delta$
in the ($m_{\tilde{g}}$, $m_{\tilde{t}_1}$) plane, for the points allowed
by the LHC
searches at the 95\%~C.L. We remind the reader that we use $\Lambda=10\tev$. 
One can see that the region with
$m_{\tilde{g}}\simeq 800\gev$
presents a large degree of fine-tuning, as could be expected since
$\mu\simeq 600\gev$ for these points. As was the case in Scenario~1, the
points with the lowest fine-tuning,
$\Delta\leq 50$, are characterized by stops masses not exceeding 800\gev,
independently
of the other parameters.
Differently from Scenario~1, however, we could not find any points
with $\Delta\leq 25$, a fact that appears clear by comparing
\reffig{fig:FT_S3}\subref{fig:b} with \reffig{fig:FT_S1}\subref{fig:a},
where the distribution of $\Delta$ for the points allowed by the LHC
is shown in the (\msusy, $m_{\neutone}$) plane for Scenarios~2 and 1, respectively.
As mentioned above, the reason is that Scenario~2 is more constrained than Scenario~1 because of the light gluinos
in the spectra.
Thus, points with low fine-tuning, which were rare in the framework of Scenario~1,
become even more difficult to find in Scenario~2.

Finally, Scenario~2, does not show relevant differences with respect
to Scenario~1 when it comes to the other phenomenological observables, 
since their values in the MSSM do not depend strongly on the gluino mass.
We found fewer points than in Scenario~1 having $\Delta\leq 50$ and being consistent with the Higgs mass measurement. 
However, the distribution on the ($\msusy$, $m_h$) plane does not look significantly different from
\reffig{fig:constr_S1}\subref{fig:a}, and we refrain from showing it over here.

The bounds from \brbsmumu\ and \brbxsgamma\ are by construction satisfied at $2\sigma$ for
the parameter space allowed by the LHC,
and the relic density assumes the same values as in Scenario~1 when $\mu$
is taken equal.
Consequently, the prospects for direct detection searches
do not change with respect to Scenario~1.


\subsection{Scenario 3}

We analyze the impact of our selected LHC searches in a more complex scenario,
whose spectra are characterized by the presence of light sleptons of the three generations,
a bino-like lightest neutralino \neutone, and wino-like \neuttwo\ and \charone, in addition to the particles of Scenario~2. 
We point out here that the level of fine-tuning in this scenario is higher than in the previous ones,
$\Delta_{\mu}\simeq100$ in Scenario~3, since the $\mu$ parameter is fixed, 
$\mu=630\gev$. We will, nonetheless, calculate the fine-tuning measure due to the other soft
SUSY-breaking parameters, hereafter indicated with $\bar{\Delta}$, to describe the impact of the contributions from the squark
and gluino sectors. 

This scenario presents some novel features. 
First, it allows investigation of the EW sector of the theory with the 
CMS 3-lepton + MET search, since the gaugino nature of \neutone, \neuttwo\ and \charone\
leads to hierarchical spectra that can produce hard leptons in the decay chain.  
Second, it allows one to investigate the impact of the ATLAS 1-lepton and CMS \alphaT\ searches
on spectra significantly more complex than the ones associated with generic SMS.
Thus, the exclusion bounds on gluinos and third generation
squarks might be altered with respect to Scenarios~1 and 2. 
Third, the gaugino nature of the neutralino leads to different dark matter signatures.

%

We do not consider in this paper the case of a wino-like
neutralino, $m_{\neutone}\approx m_{\charone}$ when $M_2<M_1$, as in that case the decay chain  
$\charone\to\neutone$ yields the same experimental features as in
the Higgsino case; i.e., the decay products are soft and
the efficiencies are very small.
Moreover, the degeneracy between the neutralino and chargino masses can in the wino dark matter case
lead to signatures of long-lived charginos\cite{Gunion:1999jr}, to which the searches 
selected for this study are not sensitive.
On the other hand, the cross section for production of \neuttwo\neutone\ and \neuttwo\charone\ 
pairs, where the heavier particle is boosted, would be highly suppressed: the former by the fact 
that the Higgsino component of both produced particles is close to zero; the latter by vanishing elements of the
neutralino mixing matrix.
Thus, in the wino neutralino case the impact of the EW sector on the bounds 
on the gluino and stop/sbottom masses would be negligible. 

\begin{figure}[t]
\centering

\label{fig:a}
\includegraphics[width=0.50\textwidth]{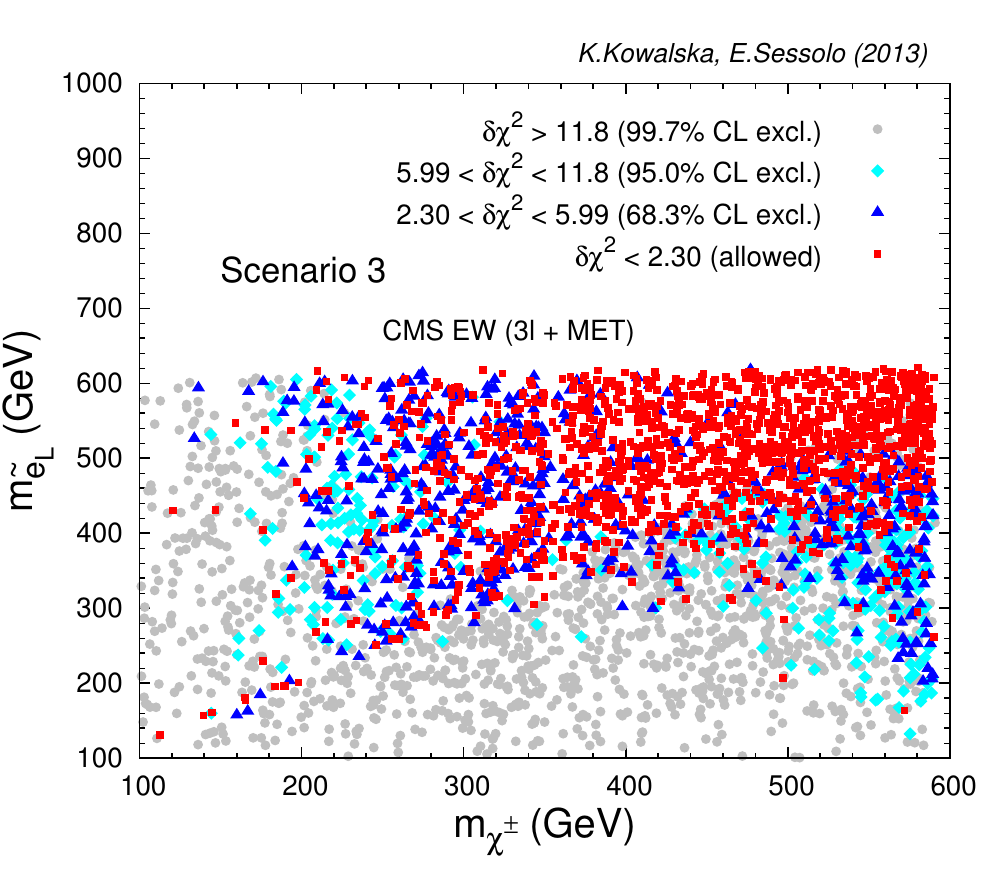}

\caption[]{\footnotesize Exclusion levels in the ($m_{\charone}$, $m_{\tilde{e}_L}$) plane from our 
simulation of the CMS 3-lepton EW production search in Scenario~3. The color code is the same as in \reffig{fig:ATLAS_21}.}
\label{fig:EW_S6}
\end{figure}

The CMS EW 3-lepton search is not sensitive to production of stops/sbottoms or gluinos, 
which yield hadronic final states with jets. In Scenario~3, it can thus only constrain 
$\tilde{\chi}_1^+\tilde{\chi}_1^-$ and \charone\neuttwo\ pair production.
For each model point the impact of this search strongly depends on whether a slepton with 
mass between the masses of \charone(\neuttwo) and \neutone\ is present in the spectrum.
Thus a scatter plot in the ($m_{\charone}$, $m_{\neutone}$) plane will look less informative than in the case of 
the SMS that we simulated for validation and comparison with the experimental result, 
shown in \reffig{fig:ATLAS_21}\subref{fig:b}.     
It is instead more instructive to look at the exclusion plot that depends on the left-handed selectron, in the
($m_{\charone}$, $m_{\tilde{e}_L}$) plane, which we show in \reffig{fig:EW_S6}.
One can identify two regions
excluded at the 95\%~C.L.: one at $m_{\charone}<200\gev$, irrespectively
of the slepton mass, where the three-body decays
$\charone\to\nu_l l^{\pm}\neutone$ and $\neuttwo\to l^+ l^-\neutone$ are mediated by
off-shell sleptons, and one at $m_{\charone}>m_{\tilde{e}_L}$, which extends to $m_{\charone}\simeq 500-600\gev$, 
where the effects of on-shell sleptons enhance the signal and increase the sensitivity. 
The sensitivity drops with increasing chargino masses,
faster for the first region since the decay
products are softer. One can also see that 
in the case of on-shell intermediate sleptons the sensitivity
bound depends strongly on the slepton mass, reaching
its maximum when $m_{\tilde{e}_L}\approx0.5m_{\charone}+0.5m_{\neutone}$, 
which is the case of the SMS shown in \reffig{fig:ATLAS_21}\subref{fig:b}. 
 
\begin{figure}[t]
\centering
\subfloat[]{
\label{fig:a}
\includegraphics[width=0.50\textwidth]{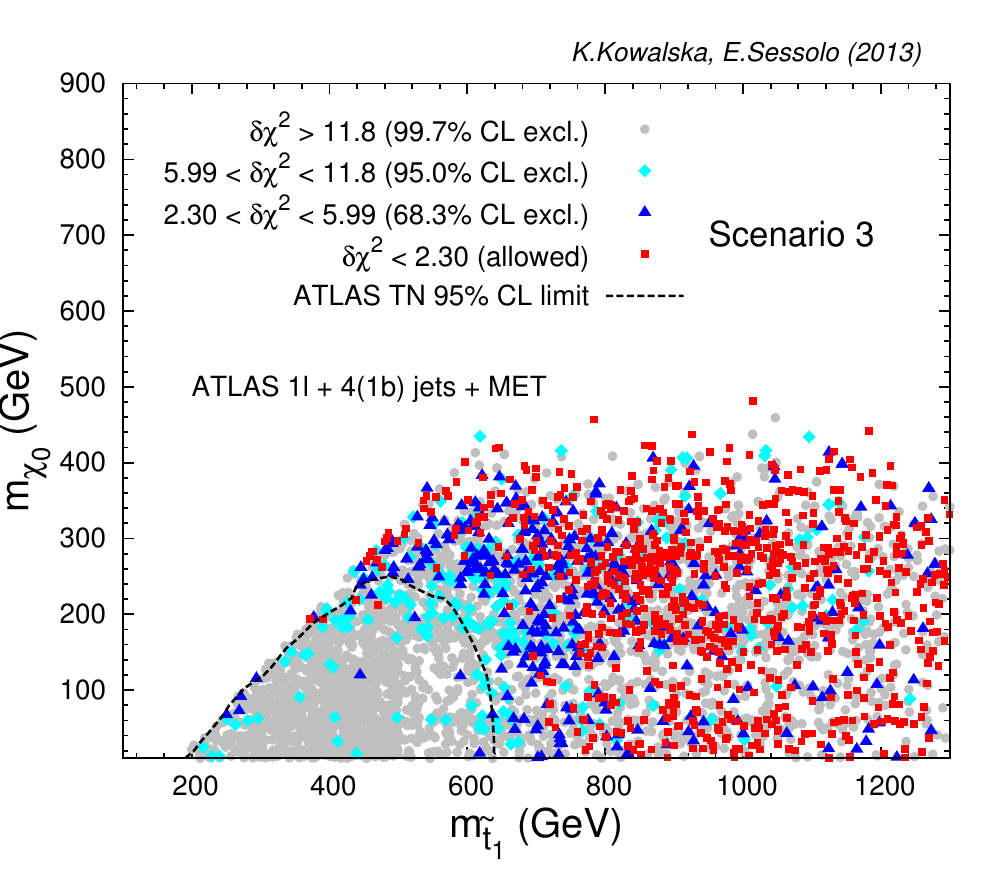}
}
\subfloat[]{
\label{fig:b}
\includegraphics[width=0.50\textwidth]{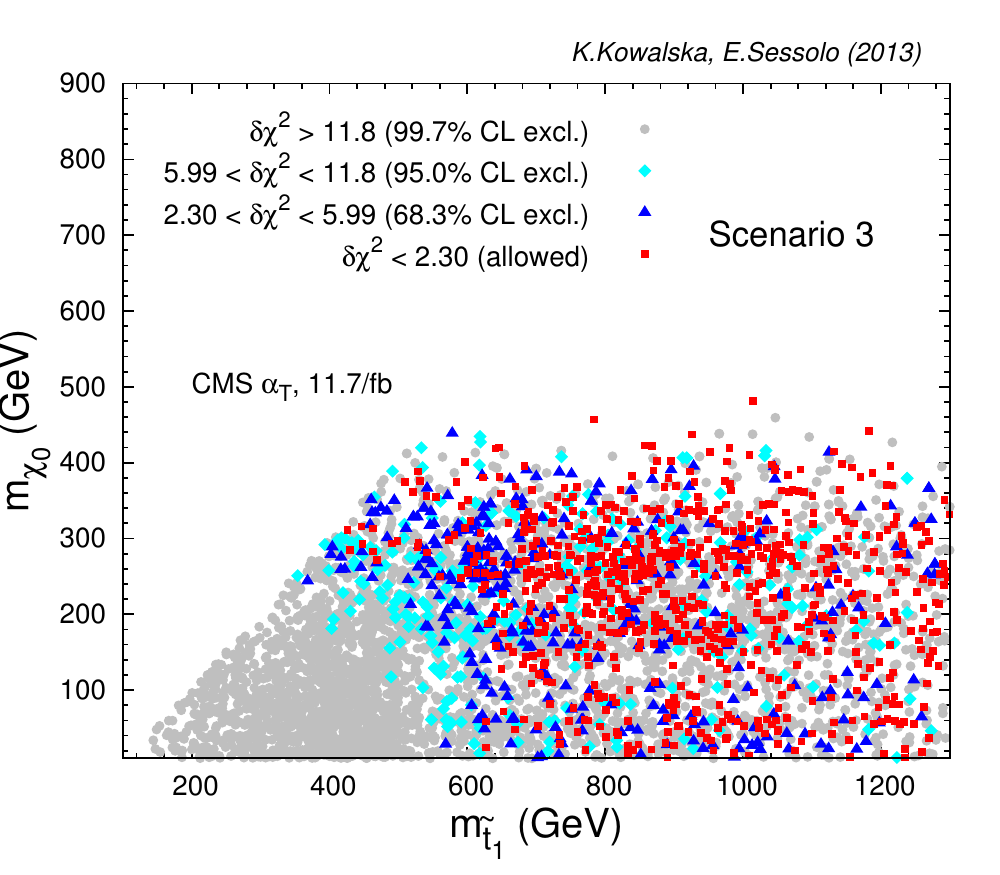}
}\\
\subfloat[]{
\label{fig:c}
\includegraphics[width=0.50\textwidth]{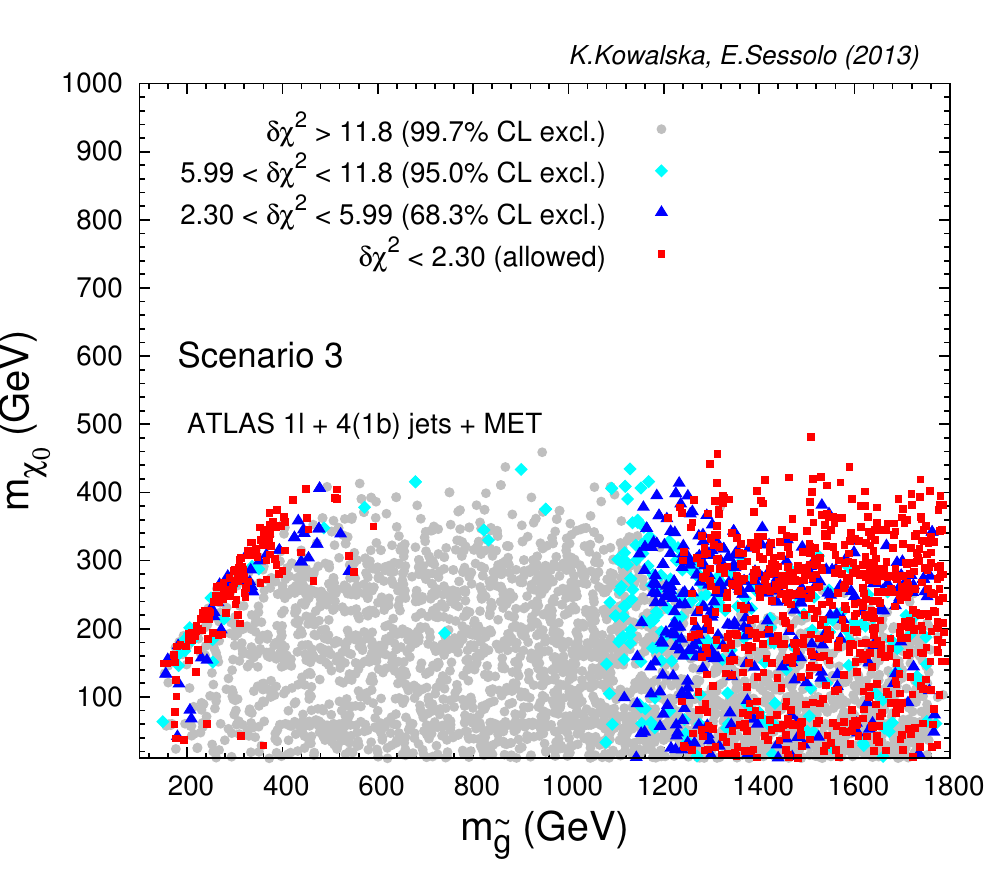}
}
\subfloat[]{
\label{fig:d}
\includegraphics[width=0.50\textwidth]{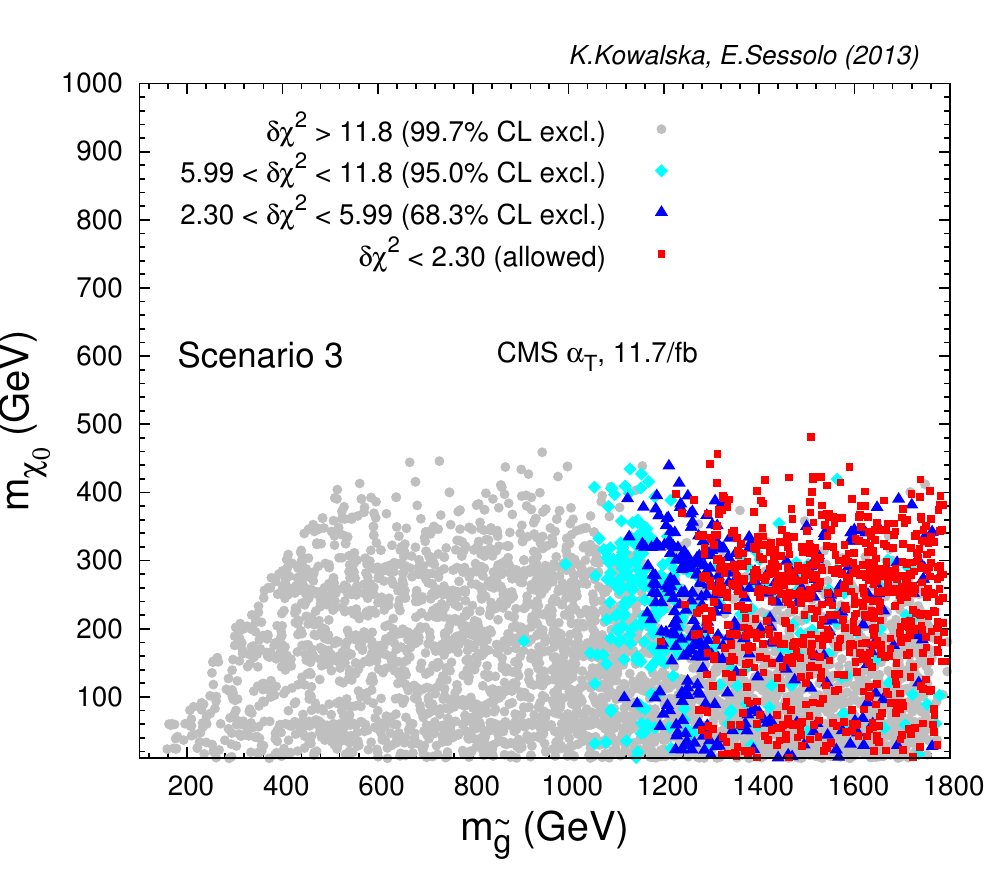}
}
\caption[]{\footnotesize Exclusion levels in the ($m_{\tilde{t}_1}$, $m_{\neutone}$) plane from our simulation of \subref{fig:a} 
the ATLAS 1-lepton search and \subref{fig:b} the CMS \alphaT\ search in Scenario~3. The dashed black line shows the published ATLAS 95\%~C.L. 
bound in SMS TN. Exclusion levels in the ($m_{\tilde{g}}$, $m_{\neutone}$) plane from our simulation of 
\subref{fig:c} the ATLAS 1-lepton search and \subref{fig:d} the CMS \alphaT\ search in Scenario~3. The color code is the same as in \reffig{fig:ATLAS_21}.}
\label{fig:COM_S6}
\end{figure}

As will appear clear below, for complex spectra it becomes very important to combine independent searches 
that investigate different experimental topologies. 
This is because, as was mentioned in \refsec{intro:sec}, the bounds on SUSY masses from an individual search can in some cases be weakened with respect to the ones obtained in the 
framework of a SMS.  
To give a practical example, we show in \reffig{fig:COM_S6}\subref{fig:a} the exclusion plot in the ($m_{\tilde{t}_1}$, $m_{\neutone}$) plane for the ATLAS 1-lepton search in Scenario~3.
One can see that, as was the case for Scenario~2, many points with $m_{\tilde{t}_1}\gg800\gev$ are excluded
due to the presence of a light gluino in the spectra. On the other hand, there are some points not excluded at the 95\%~C.L., 
or at the 99.7\%~C.L., in the region of the parameter space that was strongly excluded in SMS TN.
Some caution is required when trying to draw definite conclusions about these points,
since their number is not large. Moreover, we repeat that our criterion for exclusion is just an approximation, and carries with it some limitations.
However, taking the exclusion confidence level at face value, we gave a closer look at the PYTHIA event distribution of these points, finding that they are characterized by a large 
number of events with no hard isolated lepton in the final state,
which give no signal, or by events that involve taus in the final state, for which reconstruction is a delicate task.
A typical decay chain is, for example, $\tilde{t}\to b\tilde{\chi}_1^+$, where the
chargino decays through intermediate $\tilde{\tau}$ or $\tilde{\nu}_{\tau}$, 
$\tilde{\chi}_1^+\to\tau^+\nu_{\tau}\neutone$, and the $\tau^+$ decays hadronically.   
It is also not trivial to investigate the effects that these events have on the overall efficiencies, 
given the large number of kinematical boxes we employ in our simulation. 
But, in any case, one can see in \reffig{fig:COM_S6}\subref{fig:b} that the \alphaT\ search produces 
a more stable exclusion line in the ($m_{\tilde{t}_1}$, $m_{\neutone}$), due to the statistical combination of different final state topologies. 

We show for comparison the exclusion plots in the ($m_{\tilde{g}}$, $m_{\neutone}$) 
for the ATLAS 1-lepton and CMS \alphaT\ searches in Figs~\ref{fig:COM_S6}\subref{fig:c} and \ref{fig:COM_S6}\subref{fig:d}, respectively.

\begin{figure}[t]
\centering
\subfloat[]{
\label{fig:a}
\includegraphics[width=0.50\textwidth]{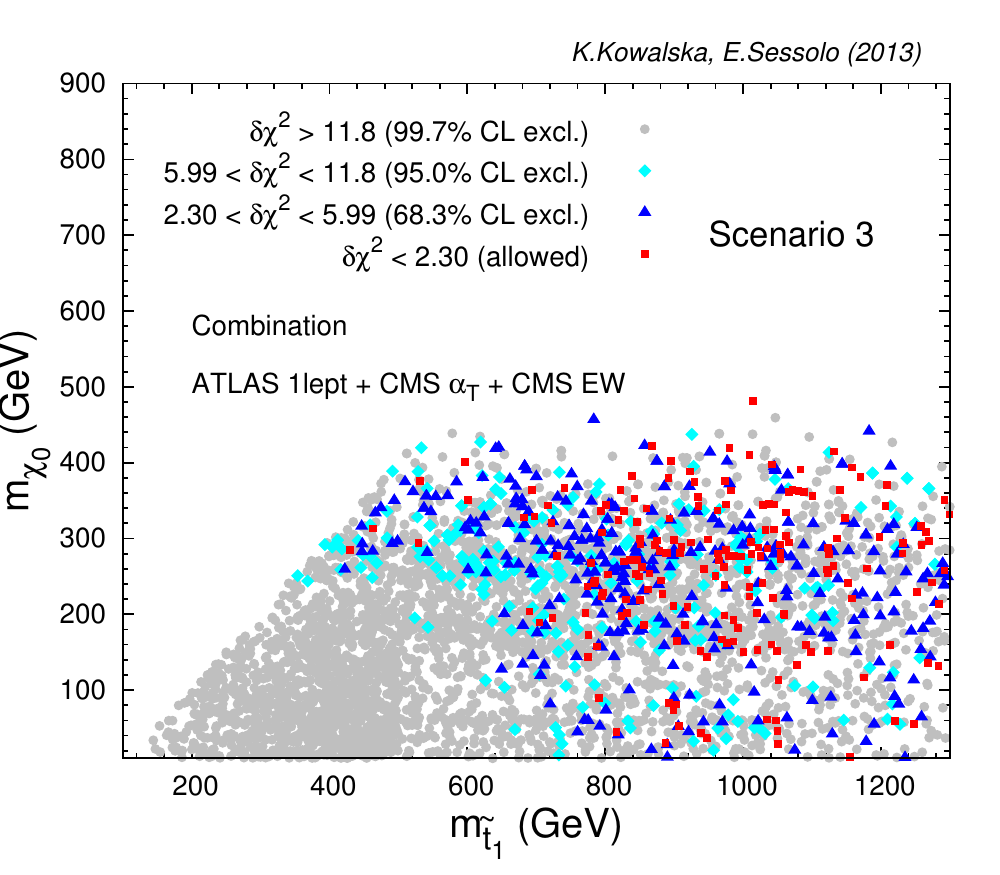}
}
\subfloat[]{
\label{fig:b}
\includegraphics[width=0.50\textwidth]{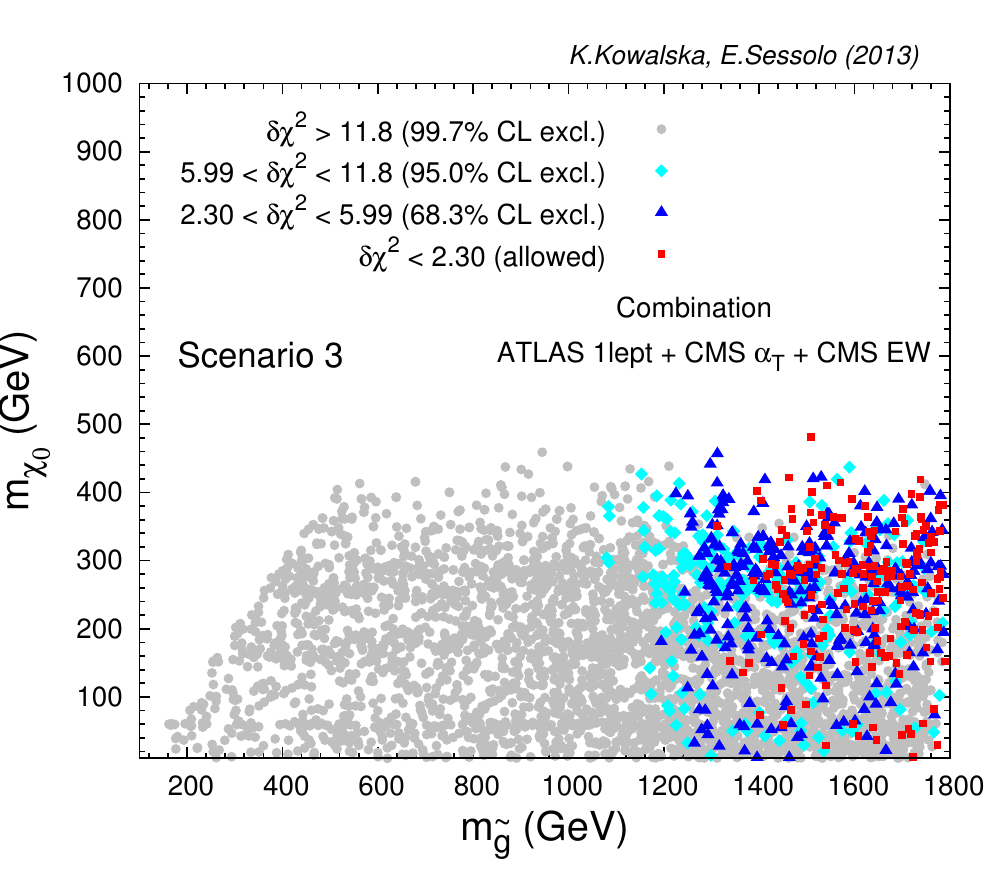}
}
\caption[]{\footnotesize  Exclusion levels from our combination of the ATLAS 1-lepton, CMS \alphaT, 
and CMS EW production searches in \subref{fig:a} the ($m_{\tilde{t}_1}$, $m_{\neutone}$) plane and \subref{fig:b} 
the ($m_{\tilde{g}}$, $m_{\neutone}$) plane in Scenario~3. The color code is the same as in \reffig{fig:ATLAS_21}.}
\label{fig:COM2_S6}
\end{figure}

In \reffig{fig:COM2_S6}\subref{fig:a} we show the statistical combination of the ATLAS 1-lepton, CMS \alphaT, 
and CMS EW production searches in the ($m_{\tilde{t}_1}$, $m_{\neutone}$) plane. In \reffig{fig:COM2_S6}\subref{fig:b} 
we show the same, in the  ($m_{\tilde{g}}$, $m_{\neutone}$) plane.
One can see that the combination of all our searches strongly reduces the number of allowed points in Scenario~3.
With the exception of a few points for which the spectra show features similar to the SMS, i.e., $m_{\tilde{t}_1}\ll m_{\tilde{g}}$ and 
\charone\ or sleptons too heavy to produce a signature in the 3-lepton search (points that become increasingly rare to find in the plots), 
\reffig{fig:COM2_S6} shows that stops are bound to $m_{\tilde{t}_1}\gsim700\gev$
for a light \neutone, while the bound on the gluino mass does not change significantly from Scenario~2.

Thus, one can see that, in spite of the limitations that might emerge with complex spectra in an individual search, 
a statistical combination of different and possibly independent searches, from both ATLAS and CMS for instance, 
stabilizes the bounds and strongly reduces the allowed regions of the parameter space, thus producing 
limits on the individual masses that are enhanced with respect to the case of selected SMS. 
Given the nature of certain decay chains observed in Scenario~3, we suspect that even stronger constraints 
might be obtained by including additional targeted searches, e.g., EW production with taus that decay hadronically in the final state\cite{ATLAS-CONF-2013-028}.

Similar conclusions were already drawn in\cite{Buchmueller:2013exa} for a combination of CMS searches at \seven,
in an original presentation style that involved ``traffic light" plots. We confirm this result over here, 
where we limit ourselves to presenting the likelihood-based exclusion levels for the points generated in our scenarios.\bigskip

\begin{figure}[t]
\centering
\subfloat[]{
\label{fig:a}
\includegraphics[width=0.50\textwidth]{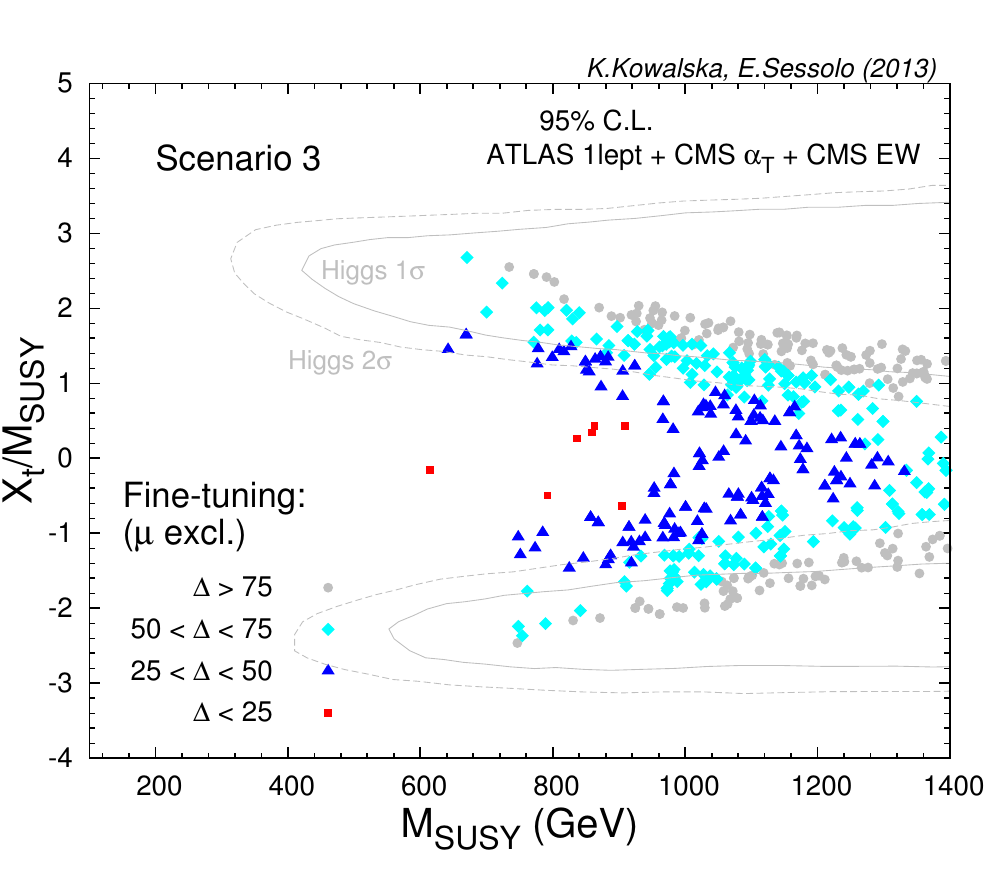}
}
\subfloat[]{
\label{fig:b}
\includegraphics[width=0.50\textwidth]{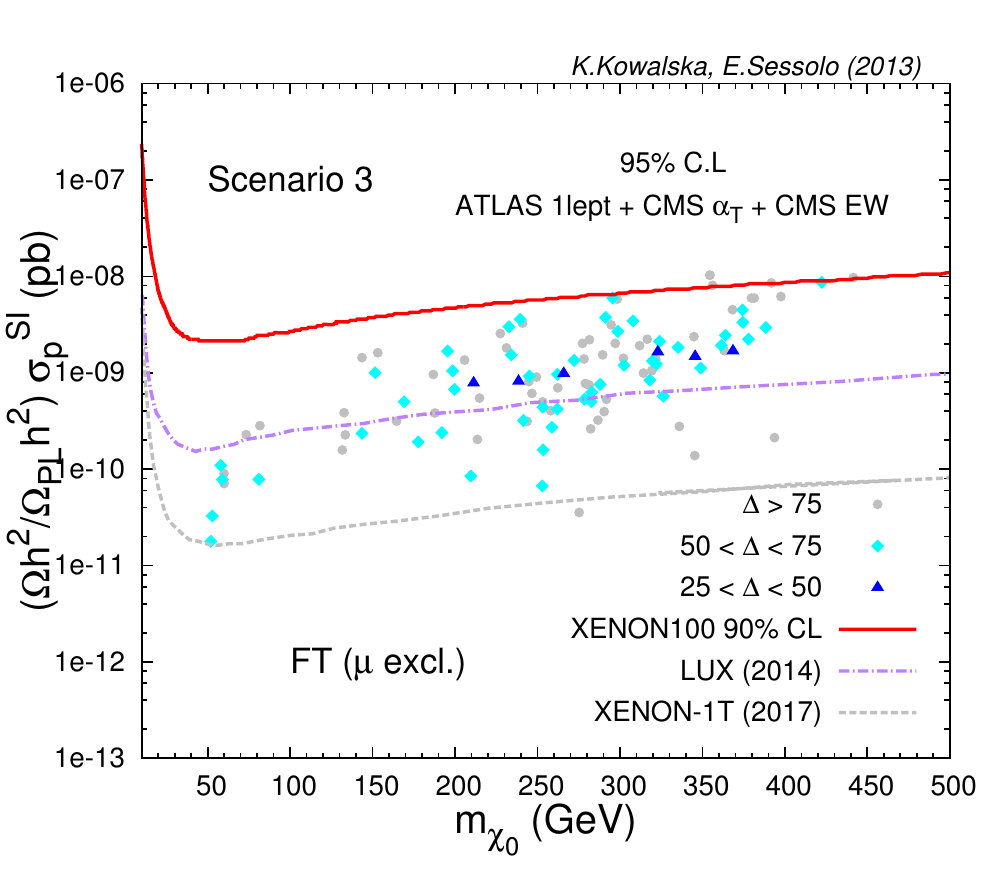}
}
\caption[]{\footnotesize \subref{fig:a} Scatter plot of the fine-tuning measure $\bar{\Delta}$ in the 
(\msusy, $X_t/\msusy$) plane for the points that are not excluded at the 95\%~C.L. by the LHC in Scenario~3.
The solid (dashed) gray contours indicate the approximate 
$1\sigma$ ($2\sigma$) window for the Higgs mass. 
\subref{fig:b} Scatter plot of $\bar{\Delta}$ in the ($m_{\neutone}$, $\Omega_{\chi} h^2/\Omega_{\textrm{Planck}}h^2\cdot\sigsip$) 
plane for the points not excluded at the 95\%~C.L by the LHC, with the constraints from the Higgs mass, relic density (upper limit), 
\brbsmumu, and \brbxsgamma\ satisfied at $2\sigma$ in Scenario~3. The solid red line shows the 90\%~C.L. bound from XENON100, 
while the dot-dashed purple and dashed gray lines show future sensitivities at LUX and XENON1T, respectively. 
The color code is the same as in \reffig{fig:FT_S1}.}
\label{fig:constr_S6}
\end{figure}

We show in \reffig{fig:constr_S6}\subref{fig:a} the distribution of the fine-tuning measure $\bar{\Delta}$ in the 
(\msusy, $X_t/\msusy$) plane for the points allowed by the LHC at the 95\%~C.L. 
We neglect the contribution due to $\mu$, $\Delta_{\mu}\simeq100$, as explained at the beginning of this subsection.
When doing so, the distribution of $\bar{\Delta}$ is entirely determined by the parameters of stop sector,
as can be inferred from the figure. We also plot in \reffig{fig:constr_S6}\subref{fig:a} the approximate $1\sigma$ (solid contours) and $2\sigma$ (dashed contours)
windows for the Higgs mass.

The distribution of $\bar{\Delta}$ shows overall lower values than the equivalent distribution for $\Delta$ in Scenario~2. 
One can see that several points with $\msusy\lesssim 900\gev$ show $\bar{\Delta}\leq 25$, whereas 
in Scenario~2 not one of the points that survived the combined LHC cuts was found with $\Delta\leq 25$, 
despite the fact that the LHC constrains Scenario~3 more strongly.
Thus, it appears to us that the greatest obstacle to obtaining MSSM spectra with an acceptable level of EW fine-tuning 
after the LHC comes from the difficulty of finding regions of the parameter space characterized by
small enough values of the parameter $\mu$. 

Clearly, inclusion of the Higgs mass constraint makes the above conclusion less relevant. 
In fact, again none of the points shown as red squares in \reffig{fig:constr_S6}\subref{fig:a} presents \mhl\ within $2\sigma$
of the experimental value (we find $\mhl\simeq 110-115\gev$ for those points). On the other hand, we find that the constraints from the signal strengths 
\rgg\ and \rzz\ have no significant impact on the points in our sample.  

Finally, the relic density shows in Scenario~3 a larger range of values than in Scenarios~1 and 2. 
However, in general bino-like neutralino dark matter tends to overclose the Universe, 
unless the $\neutone\neutone$ annihilation rate is boosted by one of the known mechanisms for obtaining the correct relic density in the MSSM; see, e.g.,\cite{Fowlie:2013oua}.  
As a matter of fact, after including the constraints from the LHC, the Higgs mass, \brbsmumu, \brbxsgamma, 
and a $2\sigma$ upper bound for the relic density, we found that only 116 points survived in our Scenario~3.

We show a scatter plot of their $\bar{\Delta}$ in the ($m_{\neutone}$, $\Omega_{\chi}h^2/\Omega_{\textrm{Planck}}h^2\cdot\sigsip$) plane in \reffig{fig:constr_S6}\subref{fig:b},
where we also show the 90\%~C.L. exclusion bound by XENON100 and the sensitivities at LUX and XENON1T.

\section{Summary}\label{sum:sec}

In this paper we investigated the impact of three different LHC direct SUSY searches on the parameter space of the MSSM, 
on which we imposed a loose requirement of naturalness, $\Delta^{-1}>1\%$ with $\Lambda=10\TeV$.  

We considered three different scenarios. In Scenario~1 the SUSY spectra consist of light stops, 
sbottoms and Higgsino-like lightest chargino and neutralino, while the other sparticles are out of reach at the LHC; 
in Scenario~2 we considered the presence of an additional light gluino in the spectra; and in Scenario~3 we considered 
a more complex kind of spectra, characterized by light stops, sbottoms, gluinos, sleptons of the three generations, a bino-like lightest neutralino and wino-like lightest chargino.    
By construction, Scenario~3 is always more fine-tuned than Scenarios~1 and 2. 

For each generated point in our scenarios we performed detailed on-the-fly simulation of the following LHC 
searches based on the \eight\ data set: the 21\invfb\ ATLAS direct stop production search
with 1 lepton in the final state, the 9.2\invfb\ CMS 3-lepton EW-production search, and the 11.7\invfb\ CMS \alphaT\ 
inclusive search for squarks and gluinos. For each point we calculated the exclusion confidence level due to the individual 
searches and to their statistical combination. We then calculated the level of fine-tuning and some relevant phenomenological observables:
the Higgs mass, Higgs signal rates, the relic density of dark matter, \brbsmumu, \brbxsgamma, and the neutralino-proton SI cross section, 
\sigsip.  

We showed that, when considering increasingly complex spectra with respect to the simplified models 
for which the experimental collaborations provide official limits on the sparticle masses, and at the same time combining different searches, 
two competing effects can emerge. 
On the one hand, more complex spectra involve longer decay chains than a SMS, which can in some occasions 
produce topologies to which an individual search is not sensitive. On the other hand, 
a combination of different searches strongly limits the available parameter space for complex, well separated spectra, 
thus overcoming the above limitations and placing strong bounds on certain scenarios.   

To give an example from our discussion, consider the region with $m_{\neutone}\lesssim250\gev$ in Scenario~3.
While it is not possible to say that stops with $600\gev\lesssim m_{\tilde{t}_1}\lesssim 700\gev$ 
are absolutely excluded by any one of our implemented searches, it is certainly more unlikely than in, 
say, Scenario~1 to find a point for which the stop mass falls in the above range and, at the same time, either gluinos 
or \charone\ and \neuttwo\ are not excluded by the remaining searches. 
We thus appreciate the effort of the experimental collaborations in providing a great number of limits obtained 
with different topologies and encourage them to produce statistical combinations of independent results, even combining the ATLAS and CMS data sets. 

As pertains to the naturalness of the scenarios considered here we showed that, if one neglects compressed spectra, 
which we did not treat in this study, the present LHC limits on the squarks of the third generation and, more importantly, 
the $\mu$ parameter exclude points with $\Delta\leq 20$. Only a handful of points in Scenario~1, characterized by $\mu\lesssim 320\gev$, 
$\msusy\lesssim 850\gev$, 
and $|A_t|\lesssim 1000\gev$, were found with $\Delta\leq 25$, and they all presented a Higgs mass well below the experimental value, 
even if one considers a large theoretical uncertainty in the Higgs mass calculation. The constraints from Higgs signal rates, \brbsmumu, and \brbxsgamma\ 
can instead be satisfied more easily for the parameter space presently allowed by the LHC.
        
As is well known, finally, for Higgsino dark matter the relic density tends to be too low with respect to the value measured by PLANCK and WMAP. 
For bino dark matter it tends instead to overclose the Universe, unless the annihilation cross section is enhanced through coannihilation or resonance effects, 
which have been largely explored in the literature. 
Nonetheless, we showed that the three scenarios considered here lie in the area of interest of direct detection experiments, 
even when rescaling their possible signal.
We presented the prospects for future observation of dark matter in these scenarios at the underground experiments LUX and XENON1T.   

\bigskip
\begin{center}
\textbf{ACKNOWLEDGMENTS}
\end{center}
  We would like to thank Maurizio Pierini for a useful e-mail exchange on the implementation of signal efficiencies. 
 We would also like to thank Leszek Roszkowski, Yue-Lin Sming Tsai, and Shoaib Munir for many discussions.
  We are funded in part by the Welcome Programme
  of the Foundation for Polish Science. 
  K.K. is supported by the EU and MSHE Grant No. POIG.02.03.00-00-013/09.
  The use of the CIS computer cluster at NCBJ is gratefully acknowledged.

\bibliographystyle{utphysmcite}	
\bibliography{myref}


\end{document}